\documentclass[aps,preprint,showpacs,showkeys,nofootinbib,tightenlines]{revtex4}

\usepackage{graphicx}  %
\usepackage{amsmath}
\usepackage{bm}  %
\usepackage{ulem}
\usepackage{slashed}

\newcommand{\bea}{\begin{eqnarray}}
\newcommand{\eea}{\end{eqnarray}}
\newcommand{\beq}{\begin{equation}}
\newcommand{\eeq}{\end{equation}}
\newcommand{\bqa}{\begin{eqnarray}}
\newcommand{\eqa}{\end{eqnarray}}

\def\mqo2{{\!\!\!}}

\begin{document}

\title{
Nonrelativistic Effective Field Theory for Axions}

\author{Eric Braaten}
\email{braaten@mps.ohio-state.edu}
\affiliation{Department of Physics,
         The Ohio State University, Columbus, OH\ 43210, USA\\}

\author{Abhishek Mohapatra}
\email{mohapatra.16@buckeyemail.osu.edu}
\affiliation{Department of Physics,
         The Ohio State University, Columbus, OH\ 43210, USA\\}

\author{Hong Zhang}
\email{zhang.5676@osu.edu}
\affiliation{Department of Physics,
         The Ohio State University, Columbus, OH\ 43210, USA\\}

\date{\today}
%\date{November 2007}

\begin{abstract}
Axions can be described by a relativistic field theory
with a real scalar field $\phi$ whose self-interaction potential  is a periodic function of $\phi$.
Low-energy axions, such as those produced in the early universe
by the vacuum misalignment mechanism,
can be described more simply by a nonrelativistic effective field theory
with a complex scalar field $\psi$ whose effective potential is a function of $\psi^*\psi$.
We determine the coefficients in the expansion of the effective potential
to fifth order in $\psi^*\psi$ by matching low-energy axion scattering amplitudes.
In order to describe a Bose-Einstein condensate of axions 
that is too dense to truncate the expansion of the effective potential in powers of  $\psi^*\psi$, 
we develop a sequence of systematically improvable approximations 
to the effective potential that  resum terms of all orders in  $\psi^*\psi$.
\end{abstract}

\smallskip
\maketitle

\section {Introduction}

The most compelling solution of the strong $CP$ problem of QCD is the {\it Peccei-Quinn mechanism},
which involves the spontaneous breaking of a $U(1)$ symmetry of a quantum field theory for
physics beyond the Standard Model \cite{Peccei:1977hh}.
This solution implies the existence of  the {\it axion},
which is the pseudo-Goldstone boson associated with the Peccei-Quinn $U(1)$ symmetry
\cite{Weinberg:1977ma,Wilczek:1977pj}.
This $U(1)$ symmetry is spontaneously broken at a scale $f_a$
called the {\it axion decay constant}.
The geometric mean of $f_a$ and the axion mass $m_a$ must be about $10^8$~eV.
Astrophysical and cosmological constraints have reduced the window of possible axion  masses 
to within one or two orders of magnitude of $10^{-4}$~eV  \cite{Kim:2008hd}.

Axions are one of the most strongly motivated possibilities for the particles 
that make up the dark matter of the universe \cite{Kim:2008hd}.
Axions can be produced in the early universe with an abundance that is compatible 
with the observed dark matter density  
by a combination of the cosmic string decay mechanism \cite{Davis:1986xc,Harari:1987ht}
and the vacuum misalignment mechanism \cite{Preskill:1982cy,Abbott:1982af,Dine:1982ah}. 
Both mechanisms produce axions that are extremely nonrelativistic.
The axions from the cosmic string decay mechanism are incoherent,
while the axions from the vacuum misalignment mechanism are coherent.
Sikivie and collaborators have pointed out that gravitational interactions  
can bring the axions in the early universe into thermal equilibrium \cite{Sikivie:2009qn,Erken:2011dz}.
Other investigators have reached similar conclusions
\cite{Saikawa:2012uk,Davidson:2013aba,Noumi:2013zga,Davidson:2014hfa}.
The thermalization of the axions can produce a Bose-Einstein condensate,
and it can drive the condensate locally towards the lowest-energy states that are accessible.

The most appropriate field-theoretic framework for axions depends on the momentum scale.
In a fundamental quantum field theory for physics beyond the Standard Model,
the axion field is the Goldstone mode of the 
complex scalar field with the Peccei-Quinn  $U(1)$ symmetry.
In a low-energy effective field theory for momentum scales small compared to $f_a$,
the axion can be represented by an elementary quantum field $\phi(x)$
that is a real Lorentz scalar.
Interactions between axions are mediated by local couplings of $\phi$ to Standard Model fields.
At smaller momentum scales below the confinement scale of QCD,
axions have local self-interactions that are generated by a potential ${\cal V}(\phi)$
that is a periodic function of $\phi$.
At still smaller momentum scales below the axion mass $m_a$,
the most appropriate framework is a nonrelativistic effective field theory
called {\it axion EFT}  \cite{Braaten:2015eeu}.
The axion is represented by an elementary quantum field $\psi(\bm{r},t)$
that is a complex scalar.
The self-interactions of axions are generated by an effective potential ${\cal V}_{\rm eff}(\psi^* \psi)$.

An important application of axion EFT is to axionic dark matter.
The nonrelativistic axions produced by the vacuum misalignment mechanism
have huge occupation numbers. They are therefore often described 
by a real-valued classical field $\phi(\bm{r},t)$ that evolves according to relativistic field equations.
However if the axions form a Bose-Einstein condensate, they 
can be described more simply by a complex-valued classical field $\psi(\bm{r},t)$ that evolves 
according to the nonrelativistic field equations of axion EFT.
Many of the theoretical issues concerning axion dark matter
can be more appropriately addressed within axion EFT.

The Lagrangian for the relativistic axion field $\phi$ is that of a real Lorentz scalar field 
with a self-interaction potential ${\cal V}(\phi)$.
One might expect that the effective Lagrangian for axion EFT could be completely determined 
from the relativistic axion  Lagrangian by a simple nonrelativistic reduction.
This is indeed the case for small fluctuations around the vacuum.
The nonrelativistic reduction of a free relativistic real scalar field $\phi$ with rest mass $m_a$
is a free nonrelativistic complex field $\psi$ with kinetic mass $m_a$. 
The relativistic equations for the real field  $\phi$ are second order in time derivatives
while the nonrelativistic equations for the complex field $\psi$ are first order in time derivatives,
so they both describe the degrees of freedom of a single particle. 
The coupling constant for $2\to 2$ axion scattering can also be obtained directly 
by a simple nonrelativistic reduction.
This determines the $(\psi^* \psi)^2$ term in the effective potential ${\cal V}_{\rm eff}(\psi^* \psi)$
for axion EFT.  However the higher powers of $\psi^* \psi$ in ${\cal V}_{\rm eff}$ 
 cannot  be determined by the simple nonrelativistic reduction.
The simplest way to determine them  is to use the matching methods of effective field theory.
Rather than deriving the effective Lagrangian from that of the relativistic theory,
its general form is assumed and the specific terms
are deduced by matching low-energy $n \to n$ axion scattering amplitudes.

In Section~\ref{sec:Relativistic},
we discuss two alternatives for the relativistic axion potential ${\cal V}(\phi)$:
the more familiar instanton potential and the more accurate chiral potential.
In Section~\ref{sec:Nonrelativistic},
we study the effective potential ${\cal V}_{\rm eff}(\psi^* \psi)$ for axion EFT.
We calculate the exact coefficients in the expansion of ${\cal V}_{\rm eff}$
to 5th order in $\psi^*\psi$ by matching low-energy axion scattering amplitudes.
If a Bose-Einstein condensate of axions is sufficiently dense, 
the expansion of ${\cal V}_{\rm eff}$ in powers of $\psi^*\psi$ cannot be truncated. 
We therefore introduce a sequence of systematically improvable approximations to ${\cal V}_{\rm eff}$
that  resum terms of all orders  in  $\psi^*\psi$.
Our results are summarized in Section~\ref{sec:Summary}.

%\newpage

\section{Relativistic axion field theory}
\label{sec:Relativistic}

In this Section,
we present two alternatives for the relativistic axion potential  ${\cal V}(\phi)$:
the instanton potential \cite{Peccei:1977ur} and the chiral potential  \cite{DiVecchia:1980yfw}.
Since the chiral potential is less familiar,
we present its derivation from the leading-order chiral Lagrangian for QCD,
following closely the analysis of Ref.~\cite{diCortona:2015ldu}.
Some properties of the axion were determined precisely
in Ref.~\cite{diCortona:2015ldu}
using a next-to-leading-order chiral Lagrangian.

\subsection{Relativistic Lagrangian for axions}

At momentum scales much smaller than the axion decay constant  $f_a$,
the axion can be described by a relativistic field theory with a real Lorentz scalar field $\phi(x)$.
At still smaller momentum scales below the QCD scale, 
the self-interactions of axions can be described by a {\it relativistic axion  potential} ${\cal V}(\phi)$.
The propagation of the axion and its self-interactions are described by the Lagrangian 
%-----------------
\begin{equation}
{\cal L} = \tfrac{1}{2}\partial_\mu \phi \partial^\mu \phi -  {\cal V}(\phi) .
\label{L-phi}
\end{equation}
%----------------- 
The corresponding Hamiltonian density is 
%-----------------
\begin{equation}
{\cal H} = \tfrac{1}{2} {\dot \phi}^2
 + \tfrac{1}{2} \nabla \phi^* \cdot \nabla \phi
+ {\cal V}(\phi) .
\label{H-phi}
\end{equation}
%----------------- 
The relativistic axion  potential ${\cal V}(\phi)$ is a periodic function of $\phi$
with period $2 \pi f_a$:
%-----------------
\begin{equation}
{\cal V}(\phi) =  {\cal V}(\phi + 2 \pi f_a) .
\label{V-periodic}
\end{equation}
%----------------- 

The relativistic potential is an even function of $\phi$,
so it can be expanded in powers of $\phi^2$.
We choose an additive constant in ${\cal V}(\phi)$  so it has a minimum of 0 at $\phi = 0$.
The quadratic term in the expansion is $\frac12 m_a^2 \phi^2$, where $m_a$ is the axion mass.
The expansion of ${\cal V}(\phi)$ to higher orders in $\phi^2$
determines the coupling constants for self-interactions of the axion.
We can define dimensionless  coupling constants $\lambda_{2n}$ 
by using the mass $m_a$ and the decay constant $f_a$ to set the scales:
%-----------------
\begin{equation}
{\cal V}(\phi) = \frac12 m_a^2 \phi^2
+ m_a^2 f_a^2 \sum_{n=2}^{\infty} \frac{\lambda_{2n}}{(2n)!} \left( \frac{\phi}{f_a} \right)^{2n}.
\label{V-series}
\end{equation}
%----------------- 
The Feynman rule for the $(2n)-$axion vertex is $-i \lambda_{2n} m_a^2/ f_a^{2n-2}$.
For most purposes, the relativistic axion field theory can be treated as a classical field theory,
because loop diagrams are suppressed by factors of $ m_a^2/f_a^2$,
which is roughly $10^{-48}$.

\subsection{Effective theory below the weak scale}

At momentum scales small compared to the masses of the $W^\pm$ and $Z^0$ bosons
but above the QCD scale,
the axion can be described by a relativistic scalar field $\phi$ whose 
self-interactions are mediated by interactions with Standard Model fields.
The terms in the effective Lagrangian that couple the axion to the Standard Model fields reduce to 
%-----------------
\begin{equation}
{\cal L}_{\rm axion} = 
\frac{1}{8\pi f_a} \left( \alpha_s G^a_{\mu \nu} \tilde G^{a\mu \nu} 
+ \frac{E}{N} \alpha F_{\mu \nu} \tilde F^{\mu \nu} \right) \phi
+ \frac{1}{2 f_a}J^\mu \partial_\mu \phi ,
\label{L-axion1}
\end{equation}
%----------------- 
%[Ref.~\cite{diCortona:2015ldu}, Eq.~(1)] 
where $G^a_{\mu \nu}$ and $F_{\mu \nu}$ are the field strengths for QCD and QED, 
$\tilde G^a_{\mu \nu} = \frac12 \epsilon_{\mu \nu \lambda \sigma} \tilde G^{a  \lambda \sigma} $ 
and $\tilde F_{\mu \nu}$ are the corresponding dual field strengths, 
and $J^\mu$ is a linear combination of axial-vector quark currents 
that depends on the details of the axion model.
The anomaly ratio $E/N$ is also model dependent.
For example, $E/N= 0$ in the simplest KSVZ model  \cite{Kim:1979if,Shifman:1979if}
and $E/N= 8/3$ in a simple DFSZ model \cite{Dine:1981rt,Zhitnitsky:1980tq}.
The QCD field-strength term in Eq.~\eqref{L-axion1}
is proportional to the topological charge density $\alpha_s G^a_{\mu \nu} \tilde G^{a\mu \nu}/8\pi$.
It defines the normalization of the axion decay constant $f_a$.
The quantization of the topological charge in the Euclidean field theory
implies a shift symmetry of the axion field, 
which requires the axion potential ${\cal V}(\phi)$ to satisfy the periodicity condition 
in Eq.~\eqref{V-periodic}.
At low momentum, the couplings of axions to the lightest quarks are particularly important.
The lightest quarks are the up and down quarks,
which form an $SU(2)$ flavor doublet:  ${\cal Q} =\binom{u}{d}$.
Their  mass term can be expressed as
%-----------------
\begin{equation}
{\cal L}_{\rm mass} = 
{\rm Tr}\big[ M ( {\cal Q}_R \bar {\cal Q}_L  + {\cal Q}_L \bar {\cal Q}_R ) \big],
\label{L-mass1}
\end{equation}
%----------------- 
where $M = {\rm diag}(m_u,m_d)$ is the mass matrix of the $u$ and $d$ quarks.

The gluon field strength term in Eq.~\eqref{L-axion1} can be eliminated 
by a chiral transformation of the quark fields that depends on $\phi(x)$.
The transformation can be restricted to the doublet ${\cal Q}$
of $u$ and $d$ quarks:
%----------------- 
\begin{equation}
{\cal Q}(x) \longrightarrow \exp\big(i (\phi(x)/2f_a) T\gamma_5\big) \, {\cal Q}(x),
\label{chiral-Q}
\end{equation}
%----------------- 
%[Ref.~\cite{diCortona:2015ldu}, Eq.~(3)] 
where $T$ is a hermitian $2 \times 2$ flavor matrix with unit trace.
The further condition that this transformation commutes with the $SU(2) \times U(1)$ 
electroweak gauge symmetry requires $T$ to be proportional to the unit matrix.
The transformation in Eq.~\eqref{chiral-Q} eliminates the QCD field-strength term in Eq.~\eqref{L-axion1},
and it  changes the coefficient of the QED field-strength term
and the coefficients of the axial-vector $u$ and $d$ currents:
%-----------------
\begin{equation}
{\cal L}_{\rm axion} = 
\frac{1}{8\pi f_a} \left( \frac{E}{N} - 6\, {\rm tr}(Q^2 T) \right) \alpha F_{\mu \nu} \tilde F^{\mu \nu}  \phi
+ \frac{1}{2 f_a}\left( J^\mu - \bar {\cal Q} T \gamma^\mu \gamma_5 {\cal Q} \right)\partial_\mu \phi, 
\label{L-axion2}
\end{equation}
%----------------- 
where $Q= {\rm diag}(+\tfrac23,- \tfrac13)$ is the charge  matrix for the $u$ and $d$ quarks.
The transformation in Eq.~\eqref{chiral-Q} changes the mass term in Eq.~\eqref{L-mass1} to
%-----------------
\begin{equation}
{\cal L}_{\rm mass} = 
{\rm Tr}\big[ {\cal M}(\phi) {\cal Q}_R \bar {\cal Q}_L  
+ {\cal M}^\dagger(\phi) {\cal Q}_L \bar {\cal Q}_R \big],
\label{L-mass2}
\end{equation}
%----------------- 
where the $2 \times 2$ matrix $ {\cal M}(\phi)$  depends on the axion field:
%-----------------
\begin{equation}
{\cal M}(\phi) = 
  \exp\big(i (\phi/2f_a) T \big) M  \exp\big(i (\phi/2f_a) T\big).
\label{Maxion}
\end{equation}
%----------------- 

\subsection{Instanton potential}

\begin{figure}[t]
\centerline{ \includegraphics*[width=12cm,clip=true]{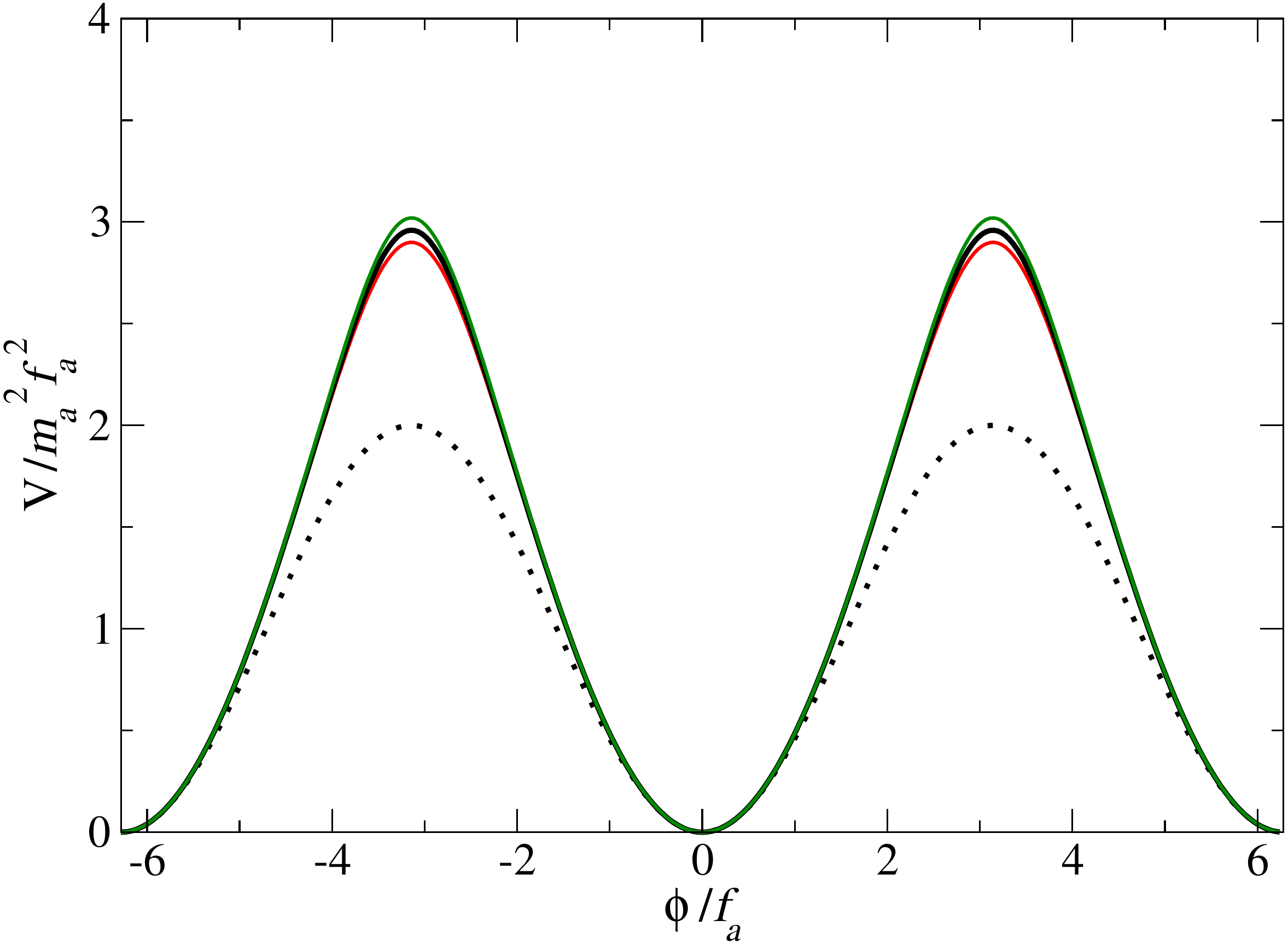} }
\vspace*{0.0cm}
\caption{Relativistic axion potentials ${\cal V}$ as functions of $\phi$:
instanton potential (dotted curve)
and chiral potential for $z=0.48$ (thick solid curve)
and for $z=0.45$ and 0.51  (higher and lower thin solid curves).
}
\label{fig:potential}
\end{figure}

The relativistic axion potential that has been used in most phenomenological studies 
of the axion is
%-----------------
\begin{equation}
{\cal V}(\phi)  =  m_a^2 f_a^2 \left[ 1 - \cos(\phi/f_a)\right] .
\label{V-instanton}
\end{equation}
%----------------- 
This potential was first derived by Peccei and Quinn \cite{Peccei:1977ur}.
We refer to it as the {\it instanton potential}. 
The dimensionless coupling constants $\lambda_{2n}$
for axion self-interactions defined by the power series in Eq.~\eqref{V-series}
are $\lambda_{2n} = (-1)^{n+1}$.
The negative sign of $\lambda_4 = -1$ implies that axion pair interactions are attractive.
The instanton potential is illustrated in Fig.~\ref{fig:potential}.

%%%%%%%%%%%%%%%%%%%%%%%%%%%%%%%%%%%%%%%%%%%%%%%%
%%%%%%%%%%%%%%%%%%%%%%%%%%%%%%%%%%%%%%%%%%%%%%%%

The instanton potential can be derived using an approximation that
keeps only terms that are leading order in the Yukawa coupling constants
and in the self-interaction coupling constants for the 
complex scalar field with the Peccei-Quinn $U(1)$ symmetry  \cite{Peccei:1977ur}. 
The instanton potential can also be derived  to all orders in the coupling constants
by using the dilute instanton gas approximation  \cite{Peccei:1977ur}.
Neither of these approximations is actually valid,
and there is no known way to systematically improve upon them.
The instanton potential should therefore be regarded at best 
as a qualitative model for the relativistic axion potential.

\subsection{Chiral potential}

QCD has an $SU(2)_L\times SU(2)_R$ symmetry that is spontaneously broken to 
its diagonal $SU(2)$ subgroup and is also explicitly broken to that subgroup 
by the quark mass term in Eq.~\eqref{L-mass1}. 
The momentum scale for the spontaneous symmetry breaking is $4 \pi f_\pi$, 
where $f_\pi \approx 92$~MeV is the pion decay constant.
At momentum scales far below $4 \pi f_\pi$,
the only degrees of freedom of QCD are the pion fields $\pi ^i(x)$, $i=1,2,3$.
They can be expressed as an $SU(2)$-valued matrix $U(x) = \exp(i \pi ^i(x) \sigma^i/f_\pi)$
whose expectation value in the QCD vacuum is the unit matrix.
The explicit symmetry breaking of the quark mass term in Eq.~\eqref{L-mass1}
can be reproduced by the pion fields by making the substitution
%-----------------
\begin{equation}
{\cal Q}_L  \bar {\cal Q}_R \longrightarrow
\frac{m_\pi^2 f_\pi^2}{2(m_u + m_d)} U(x).
\label{Q-U}
\end{equation}
%----------------- 
where $m_\pi \approx 140$~MeV is the pion mass.
The quark mass term in Eq.~\eqref{L-mass2} then becomes
%-----------------
\begin{equation}
{\cal L}_{\rm mass} = 
 \frac{m_\pi^2 f_\pi^2}{2(m_u + m_d)} 
 {\rm Tr}\big[ {\cal M}(\phi) U^\dagger + {\cal M}^\dagger(\phi) U \big].
\label{L-mass3}
\end{equation}
%----------------- 

One might be tempted to obtain the axion self-interaction potential ${\cal V}(\phi)$ 
by taking the expectation value of ${\cal L}_{\rm mass}$ in Eq.~\eqref{L-mass3}
in the QCD vacuum, which can be obtained by setting $U$ equal to the identity matrix.
It is evident that this is incorrect, because 
$ {\rm Tr}[ {\cal M}(\phi)  + {\cal M}^\dagger(\phi)]$
does not satisfy the periodicity condition in Eq.~\eqref{V-periodic},
and it depends on the choice of the flavor matrix $T$ in the chiral transformation in Eq.~\eqref{chiral-Q}.
To obtain the correct axion potential, it is necessary to take into account the
response of the pion fields.  This can be accomplished by 
setting the pion fields  in Eq.~\eqref{L-mass3}
equal to their stationary values in the presence of a constant axion field $\phi$.
The resulting potential for the axion field is
%-----------------
\begin{equation}
{\cal V}(\phi) =
m_\pi^2 f_\pi^2 \left(1- \left[ 1- \frac{4 z}{(1+z)^2} \sin^2(\phi/2f_a) \right]^{1/2} \right),
\label{V-chiral}
\end{equation}
%----------------- 
where $z = m_u/m_d$ is the ratio of the up and down quark masses.  
This potential was first derived by  Di Vecchia and Veneziano \cite{DiVecchia:1980yfw}.
We refer to it as the {\it chiral potential}. 
The product of the axion mass and the axion decay constant is given by
%-----------------
\begin{equation}
m_a^2  f_a^2 = \frac{z}{(1+z)^2} m_\pi^2 f_\pi^2 .
\label{mafa-pi}
\end{equation}
%----------------- 

In Ref.~\cite{diCortona:2015ldu}, the analysis of the axion potential was carried out to
next-to-leading order in the chiral effective field theory for QCD. 
The numerical value of the up/down quark mass ratio  is $z = 0.48(3)$.
The product of the axion mass and decay constant is  \cite{diCortona:2015ldu}
%-----------------
\begin{equation}
m_a  f_a = \left[ 7.55(5) \times 10^7~{\rm eV} \right]^2.
\label{mafa}
\end{equation}
%----------------- 
Cosmological constraints restrict the decay constant to $f_a \lesssim 10^{21}~{\rm eV}$,
and astrophysical constraints restrict it to $f_a \gtrsim 3\times 10^{18}~{\rm eV}$  \cite{Kim:2008hd}.
The allowed range for the axion mass is therefore
$6 \times 10^{-6}~{\rm eV} \lesssim m_a \lesssim 2 \times10^{-3}~{\rm eV}$.

In Fig.~\ref{fig:potential}, the chiral potential
is compared to the instanton potential in Eq.~\eqref{V-instanton}.
The potentials have the same curvature near the minima, but the amplitude of the oscillation
for the chiral potential with $z = 0.48(3)$ is larger by a factor of 1.48(3). 
Note that the instanton potential can be derived from the chiral potential in Eq.~\eqref{V-chiral}
by applying the binomial expansion to the square root,
truncating the expansion after the $\sin^2(\phi/2f_a)$ term,
and then using a trigonometric identity.

We now compare the predictions of the chiral potential in Eq.~\eqref{V-chiral} 
and the instanton potential  in Eq.~\eqref{V-instanton}
for the dimensionless coupling constants $\lambda_{2n}$
for axion self-interactions defined by the power series in Eq.~\eqref{V-series}.
For the instanton potential, these  coupling constants  are $\lambda_{2n} = (-1)^{n+1}$.
For the chiral potential, there is no analytic expression for $\lambda_{2n}$ as a function of $n$.
The dimensionless coupling constant for the $4-$axion vertex is
%-----------------
\begin{equation}
\lambda_4 = - \frac{1-z+z^2}{(1+z)^2}.
\label{lambda4}
\end{equation}
%----------------- 
For $z=0.48(3)$, this coupling constant is $\lambda_4 = -0.343(15)$. 
The negative sign implies that axion pair interactions are attractive,
but their amplitude is smaller than for the instanton potential by about a factor of 3.
For the  chiral potential with $z=0.48$, the next three dimensionless coupling constants are 
$\lambda_6 = -0.126$, $\lambda_8 = -0.874$, and  $\lambda_{10} = -5.63$.
For the instanton potential, the coupling constants alternate in sign.
For the chiral potential with $z=0.48$, the first 7 coupling constants are negative.
They are followed by 7 positive coefficients, then 7 negative coefficients.
This pattern seems to continue indefinitely, 
although occasionally there are only 6 consecutive same-sign coupling constants.

%%%%%%%%%%%%%%%%%%%%%%%%%%%%%%%%%%%%%%%%%%%%%%%%
\begin{figure}[t]
\centerline{\large \hspace{1.5cm} Instanton \hspace{5cm} ~~Chiral ($z=0.48$)}
 \includegraphics[scale=0.45]{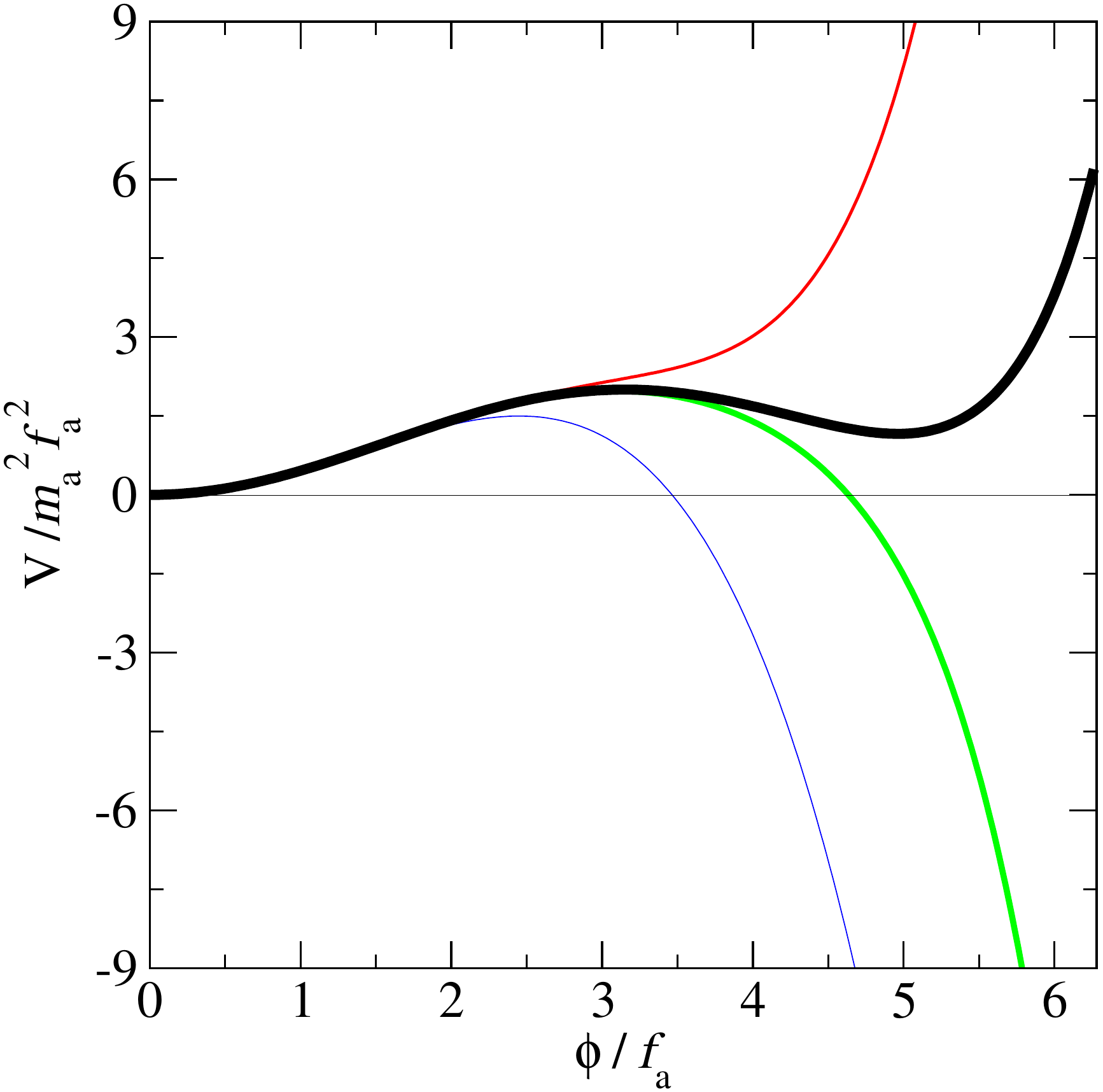}
\hspace{0.2cm}
 \includegraphics[scale=0.45]{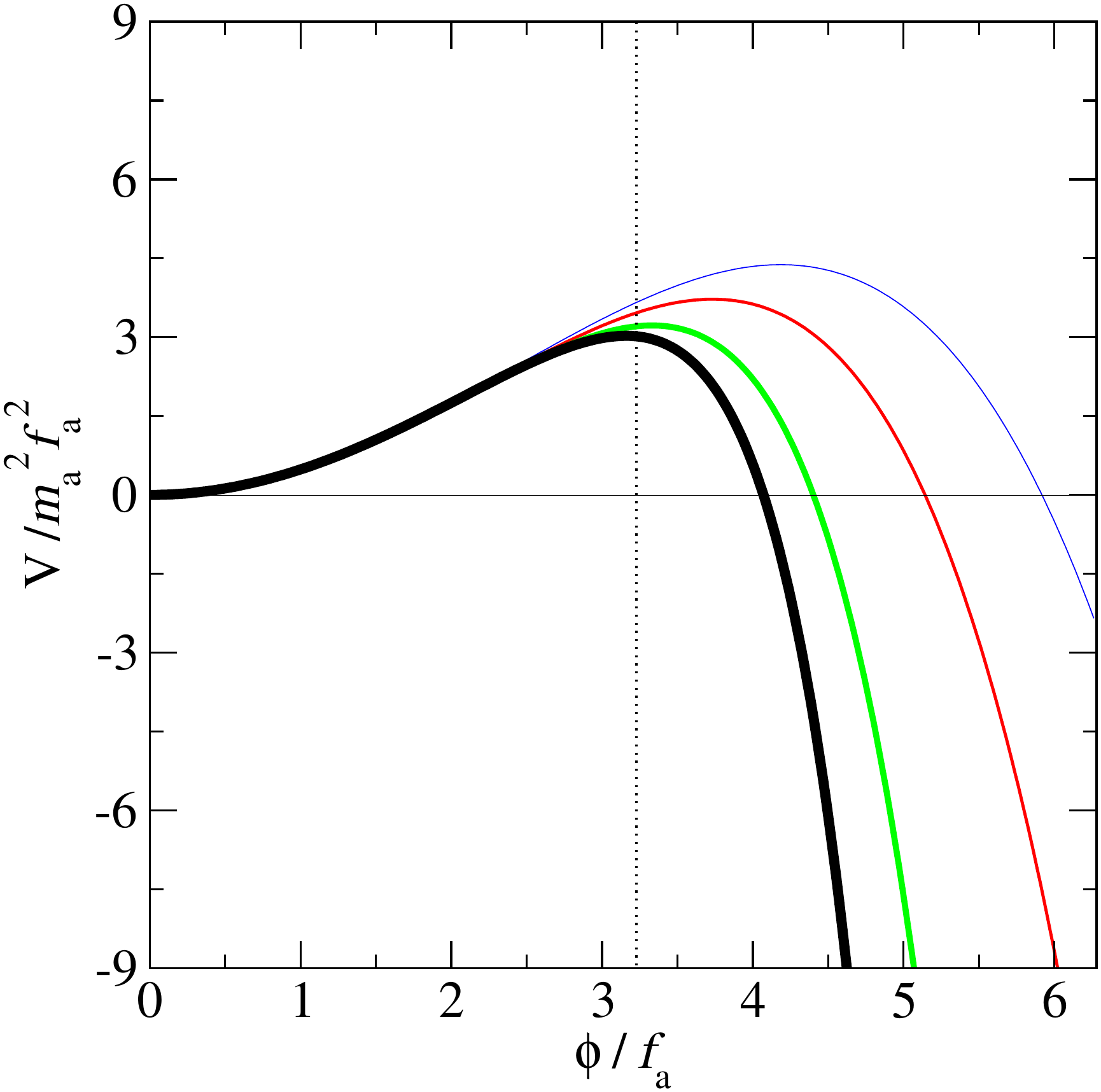}
\vspace*{0.0cm}
\caption{
Polynomial truncations of the potentials ${\cal V}$  as functions of $\phi$:
instanton potential (left panel)
and chiral potential with $z = 0.48$ (right panel).
The truncations of the potential after the 2nd, 3rd, 4th, and 5th powers of $\phi^2$
are shown as successively thicker solid lines.
The vertical dotted line in the right panel
marks the radius of convergence of the chiral  potential.
}
\label{fig:V5}
\end{figure}
%%%%%%%%%%%%%%%%%%%%%%%%%%%%%%%%%%%%%%%%%%%%%%%%

There is an important difference between the chiral potential and the instanton potential 
in the convergence properties of the power series in Eq.~\eqref{V-series}.
For the instanton potential in Eq.~\eqref{V-instanton},
the power series has an infinite radius of convergence.
For the chiral potential in Eq.~\eqref{V-chiral}, the radius of convergence is 
determined by the branch point of the square root that is closest to the origin.
The radius of convergence in $\phi/f_a$ is 
%-----------------
\begin{equation}
r_c = 2|\arcsin[(1+z)/(2z^{1/2})]| .
\label{r-conv}
\end{equation}
%----------------- 
For $z = 0.48(3)$, the radius of convergence 3.226(15) 
is a little beyond the first maxima in the potential
in Fig.~\ref{fig:potential}, which are at $\phi/f_a = \pm \pi$. 
The difference between the convergence properties of the power series
for the chiral potential and the instanton potential can be seen already at relatively
low orders in $\phi$, as illustrated in Fig.~\ref{fig:V5}.
For the instanton potential, as the order of the truncation increases,
the range of $\phi$ over which the difference between two successive approximations
is small gets larger and larger.
For the chiral potential, as the order of the truncation increases,
the range of $\phi$ over which the difference between two successive approximations
is small never gets larger than about $r_c f_a$.

Since the relativistic potential has the periodicity property in Eq.~\eqref{V-periodic},
it can be expanded in a cosine series:
%-----------------
\begin{equation}
{\cal V}(\phi) = 
m_a^2 f_a^2  \sum_{j=1}^\infty b_j \left[ \cos(j\phi/f_a) -1 \right] .
\label{V-cosine}
\end{equation}
%----------------- 
For the instanton potential in Eq.~\eqref{V-instanton}, this series has only the $j=1$ term.
For the chiral potential in Eq.~\eqref{V-chiral}, 
the coefficients $b_j$ in the cosine series 
can be expressed in terms of hypergeometric functions:
%-----------------
\begin{equation}
b_j=
\frac{(-1)^j (2j-2)!  (1+z) z^{j-1}}{2^{2j-2} j!\, (j-1)!\, (1+z^2)^{j-1/2}}\,
{}_2F_1\big( \tfrac12j - \tfrac14, \tfrac12j + \tfrac14; j+1;\big[\tfrac{2z}{1+z^2}\big]^2 \big) .
\label{bj-chiral}
\end{equation}
%----------------- 
These coefficients satisfy
%-----------------
\begin{equation}
\sum_{j=1}^\infty j^2 b_j = -1\,,
\label{bj-chiral:sum}
\end{equation}
%-----------------
which ensures that the potential in Eq.~\eqref{V-cosine} has the correct mass term $\frac12 m_a^2\phi^2$.
Using the asymptotic behavior of the hypergeometric function for large $j$ \cite{Watson:1918},
we can determine the asymptotic behavior of the coefficients:
%-----------------
\begin{equation}
b_j \longrightarrow 
\frac{(-1)^j   (1+z) (1-z^2)^{1/2} z^{j-1}}{\sqrt{\pi}\,  j^{3/2}}\,.
\label{bj-chiral:largej}
\end{equation}
%----------------- 
As $j$ increases, these coefficients decrease exponentially as $z^j$.
Thus the sum over $j$ in Eq.~\eqref{V-cosine} converges uniformly and rapidly.
A relatively low-order truncation of the cosine series gives an accurate approximation
to the chiral potential.
We can ensure that the coefficient of the mass term $\frac12 m_a^2 \phi^2$ is exact
by not using the expression in Eq.~\eqref{bj-chiral} for the last coefficient $b_{j_{\rm max}}$, 
but instead setting it equal to
%-----------------
\begin{equation}
b_{j_{\rm max}} = - \frac{1}{j_{\rm max}^2}\left(1 + \sum_{j=1}^{j_{\rm max}-1} j^2 b_j \right).
\label{bj-max}
\end{equation}
%----------------- 
For $z = 0.48$, the truncation with $j_{\rm max} = 5$ gives an error that is less than 
$10^{-3}$ of the peak value ${\cal V}_{\rm max}$ of the potential. 
The coefficients are $b_1= -1.436$, $b_2= 0.167$, $b_3= -0.0394$, and $b_4= 0.0117$
from Eq.~\eqref{bj-chiral} and $b_5= -0.0026$ from Eq.~\eqref{bj-max}.
The error decreases to $10^{-5}$ of ${\cal V}_{\rm max}$ for $j_{\rm max} = 8$. 

The cosine series in Eq.~\eqref{V-cosine} for the chiral potential 
can be used to develop a resummation method for the power series
in Eq.~\eqref{V-series} when $\phi$ is outside the radius of convergence.
The coefficient $\lambda_{2n}$ in the power series can be expressed as an infinite sum:
%-----------------
\begin{equation}
\lambda_{2n} = 
\sum_{j=1}^\infty(-1)^n j^{2n} b_j .
\label{lambda2n-bj}
\end{equation}
%----------------- 
The cosine function $\cos(j \phi/f_a)$ has a power series with an infinite radius of convergence.
The truncated cosine expansion defined by truncating the sum over $j$ 
in Eq.~\eqref{V-cosine} after the $j_{\rm max}$ term 
therefore also has a power series with an infinite radius of convergence.
The coefficient $\lambda_{2n}(j_{\rm max})$ for that power series
is obtained by truncating the sum in Eq.~\eqref{lambda2n-bj} after the $j_{\rm max}$ term.
It converges to $\lambda_{2n}$ as $j_{\rm max} \to \infty$.
As $j_{\rm max}$ is increased, the power series for the truncated cosine expansion 
converges  as a function of $j_{\rm max}$ to ${\cal V}(\phi)$ out to increasingly larger values of $\phi$.

%\newpage

\section{Nonrelativistic effective field theory}
\label{sec:Nonrelativistic}

In this section, we use effective field theory to obtain systematically improvable approximations 
to the effective potential ${\cal V}_{\rm eff}(\psi^* \psi)$ for axion EFT.
We calculate the first five coupling constants in the 
expansion of ${\cal V}_{\rm eff}$ in powers of $\psi^* \psi$
exactly by matching low-energy axion scattering amplitudes.
We also introduce a systematically improvable sequence of approximations for ${\cal V}_{\rm eff}$
 in which terms of all orders in $\psi^* \psi$ are resummed.

\subsection {Effective Lagrangian for axion EFT}

At momentum scales much smaller than the axion mass  $m_a$,
the axion can be described by a 
nonrelativistic effective field theory with a complex scalar field $\psi(\bm{r},t)$.
We refer to this effective field theory as {\it axion EFT} \cite{Braaten:2015eeu}. 
The self-interactions of axions are described by an {\it effective potential} ${\cal V}_{\rm eff}(\psi^* \psi) $.
The propagation of the axion and its self-interactions 
are described by the effective Lagrangian
%-----------------
\begin{equation}
{\cal L}_{\rm eff} = 
\tfrac12 i \left( \psi^* \dot \psi - \dot \psi^*\psi \right) - {\cal H}_{\rm eff},
\label{Leff-psi}
\end{equation}
%----------------- 
where the effective Hamiltonian density has the form
%-----------------
\begin{equation}
{\cal H}_{\rm eff} = 
\frac{1}{2m_a} \nabla \psi^* \cdot \nabla \psi
+ {\cal V}_{\rm eff}(\psi^* \psi) .
\label{Heff-psi}
\end{equation}
%----------------- 
The periodicity of $ {\cal V}(\phi)$ in Eq.~\eqref{V-periodic} 
does not impose any simple constraints on 
${\cal V}_{\rm eff}(\psi^* \psi)$.  Instead it implies that the same effective potential
${\cal V}_{\rm eff}(\psi^* \psi)$ describes the nonrelativistic field theory
associated with fluctuations of $\phi$ around
any of the minima $n(2 \pi f_a)$  of $ {\cal V}(\phi)$, where $n$ is an integer.

The effective potential for axion EFT can be expanded in powers of $\psi^* \psi$.
We choose an additive constant in ${\cal V}_{\rm eff}$  so it has a minimum of 0 at $\psi^* \psi = 0$.
The first term in the expansion is $m_a \psi^* \psi$, where $m_a$ is the axion mass.
The expansion of  ${\cal V}_{\rm eff}$ to higher orders in $\psi^* \psi$
determines the coupling constants for self-interactions of the axion.
We define dimensionless  coupling constants $v_{n}$ 
by using the mass $m_a$ and the decay constant $f_a$ to set the scales:
%-----------------
\begin{equation}
{\cal V}_{\rm eff}(\psi^* \psi) = m_a \psi^* \psi
+ m_a^2 f_a^2 \sum_{n=2}^{\infty} \frac{v_n}{(n!)^2} \left(\frac{\psi^* \psi}{2m_af_a^2} \right)^n.
\label{Veff-series}
\end{equation}
%----------------- 
The Feynman rule for the $n \to n$ axion vertex is $-i v_n m_a^2 f_a^2 / (2m_af_a^2)^n$.

The Lagrangian for axion EFT in Eq.~\eqref{Leff-psi} has a $U(1)$ symmetry
in which the field $\psi(\bm{r},t)$ is multiplied by a phase.
This symmetry implies  conservation of the axion number: 
%-----------------
\begin{equation}
N = \int\!\! d^3r\,  \psi^* \psi .
\label{Naxion}
\end{equation}
%----------------- 
In contrast, the number of axions is not conserved in the relativistic theory.
For example, $2p$ low-energy axions with $p\ge 2$ can scatter into
2 relativistic axions through the $(2p+2)-$axion vertex.
This reaction cannot be described explicitly within axion EFT, because the final-state axions are relativistic.
By the optical theorem, the rate for this reaction is proportional to the imaginary part 
of a one-loop $2p \to 2p$ scattering amplitude.
The effects of this reaction on low-energy axions can therefore be reproduced in axion EFT 
by a $(\psi^* \psi)^{2p}$ term in the Lagrangian with an imaginary coefficient.
Since this term comes from a one-loop diagram in the relativistic theory,
its coefficient is  suppressed relative to the coefficient of the $n=2p$ term 
in Eq.~\eqref{Veff-series} by a factor of $m_a^2/f_a^2$, which is roughly $10^{-48}$.
We will therefore ignore the imaginary part of the effective potential.

\subsection{Coefficients in the effective potential}
\label{sec:EFT}

The effective potential ${\cal V}_{\rm eff}$ 
for axion EFT can be derived using the matching methods of effective field theory.
One assumes that low-energy axions
can be described equally well by the Lagrangian for the relativistic real scalar field $\phi$
in Eq.~\eqref{L-phi} or
by the Hamiltonian for the nonrelativistic complex field $\psi$ of axion EFT in Eq.~\eqref{Heff-psi}.
The effective potential ${\cal V}_{\rm eff}$ is then determined by matching low-energy 
scattering amplitudes in the relativistic theory and in axion EFT.
Since loop diagrams in the relativistic theory are suppressed by factors of $m_a^2/f_a^2$,
it is only necessary to match the contributions to low-energy scattering amplitudes 
from tree-level diagrams in the relativistic theory and tree-level diagrams in axion EFT.

The matching procedure of effective field theory determines the  power series in $\psi^* \psi$
for the effective potential in Eq.~\eqref{Veff-series}.  In the mass term 
$m_a \psi^*\psi$, the coefficient is determined by the axion mass.
The coefficients of the higher powers of $\psi^* \psi$ can be determined 
by matching low-energy scattering amplitudes.
We begin by matching $2 \to 2$ axion scattering amplitudes.
The only tree-level diagram for $2 \to 2$ scattering is the $2 \to 2$ vertex.
The Feynman rule for that vertex in the relativistic theory is $-i \lambda_4m_a^2/f_a^2$.
The Feynman rule for that vertex in axion EFT is $-i v_2/4f_a^2$.
To obtain the scattering amplitude with the
standard relativistic normalization of single-particle states,
this must be multiplied by four factors of $(2 m_a)^{1/2}$.
By matching the scattering amplitudes, we obtain $v_2 = \lambda_4$.

%%%%%%%%%%%%%%%%%%%%%%%%%%%%%%%%%%%%%%%%%%%%%%%%
\begin{figure}[t]
\centerline{ \includegraphics*[width=12cm,clip=true]{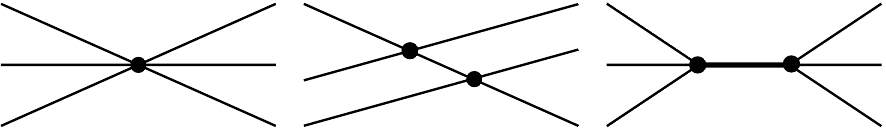} }
\vspace*{0.0cm}
\caption{The tree-level diagrams for low-energy $3 \to 3$ scattering in the relativistic axion theory.
The first 2 diagrams are also diagrams in axion EFT. 
In the last diagram, the thicker line indicates a virtual axion whose invariant mass 
is approximately $3m$. 
}
\label{fig:3to3tree}
\end{figure}
%%%%%%%%%%%%%%%%%%%%%%%%%%%%%%%%%%%%%%%%%%%%%%%%

We proceed to match the $3 \to 3$ axion scattering amplitudes.
The tree-level diagrams for $3 \to 3$ axion scattering in the relativistic theory 
are shown in Fig.~\ref{fig:3to3tree}.
The first diagram is the $3 \to 3$ vertex.
The second diagram has two $2 \to 2$ vertices connected by a virtual axion line.
These two diagrams also give contributions to $3 \to 3$  scattering in axion EFT.
The third diagram has 3 axions scattering into a single virtual axion
and then back into three axions.
This diagram does not contribute in axion EFT, 
because the invariant mass of the virtual axion is approximately $3m$.
Having already determined $v_2$,
we determine $v_3$ by matching the $3 \to 3$ axion scattering amplitudes:
$v_3 = \lambda_6  - (17/8) \lambda_4^2$.

%%%%%%%%%%%%%%%%%%%%%%%%%%%%%%%%%%%%%%%%%%%%%%%%
\begin{figure}[t]
\centerline{ \includegraphics*[width=16cm,clip=true]{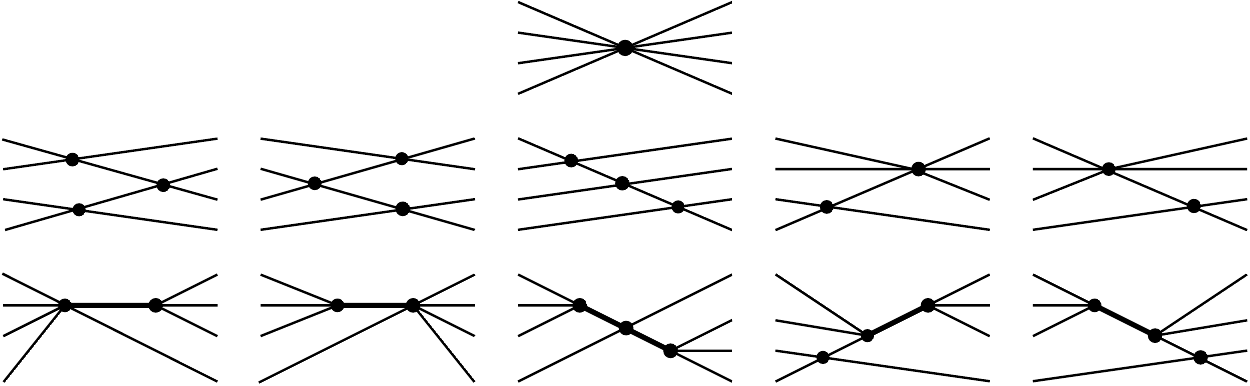} }
\vspace*{0.0cm}
\caption{The tree-level diagrams for low-energy $4 \to 4$ scattering in the relativistic axion theory.
The first 6 diagrams are also diagrams in axion EFT. 
In the last 5 diagrams, the thicker lines indicate virtual axions whose invariant mass 
is approximately $3m$. 
}
\label{fig:4to4tree}
\end{figure}
%%%%%%%%%%%%%%%%%%%%%%%%%%%%%%%%%%%%%%%%%%%%%%%%

We can determine $v_4$ by matching the $4 \to 4$ scattering amplitudes
from the tree-level diagrams in  Fig.~\ref{fig:4to4tree}.
There are 11 diagrams, 6 of which also contribute in axion EFT.
We can then determine $v_5$ by matching the $5 \to 5$ scattering amplitudes
from the tree-level diagrams.
There are 48 diagrams, 17 of which also contribute in axion EFT.
The diagrams have up to 3 virtual axion lines,
some with invariant mass that is approximately $5m$.

The exact results for the dimensionless coupling constants $v_n$
in the effective potential for axion EFT in Eq.~\eqref{Veff-series}
from matching the $n \to n$ scattering amplitudes with $n$ up to 5 are
%-----------------
\begin{subequations}
\begin{eqnarray}
v_2 &=& \lambda_4,
\label{v2}
\\
v_3 &=&  \lambda_6  -\frac{17}{8} \lambda_4^2,
\label{v3}
\\
v_4 &=&  \lambda_8 - 11 \lambda_4 \lambda_6 + \frac{49}{4} \lambda_4^3,
\label{v4}
\\
v_5 &=&  \lambda_{10} - \frac{45}{2} \lambda_4  \lambda_8 - \frac{131}{6} \lambda_6^2 
+ \frac{4315}{24} \lambda_4^2  \lambda_6 - \frac{51725}{384} \lambda_4^4. 
\label{v5}
\end{eqnarray}
\label{v2345}%
\end{subequations}
%----------------- 

%%%%%%%%%%%%%%%%%%%%%%%%%%%%%%%%%%%%%%%%%%%%%%%%
\begin{figure}[t]
\centerline{\large \hspace{1.5cm} Instanton \hspace{5cm} ~~Chiral ($z=0.48$)}
 \includegraphics[scale=0.45]{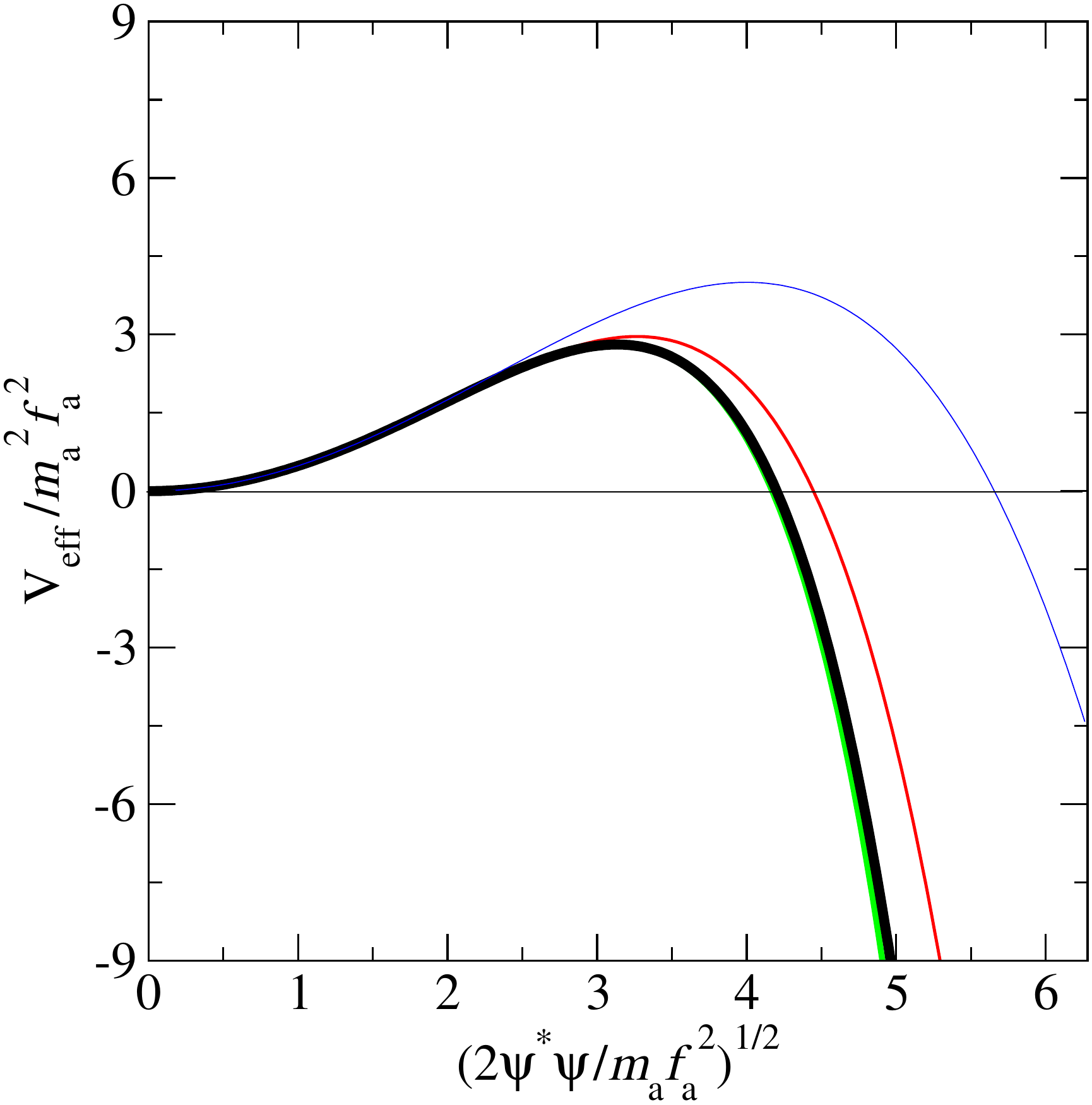}
\hspace{0.2cm}
 \includegraphics[scale=0.45]{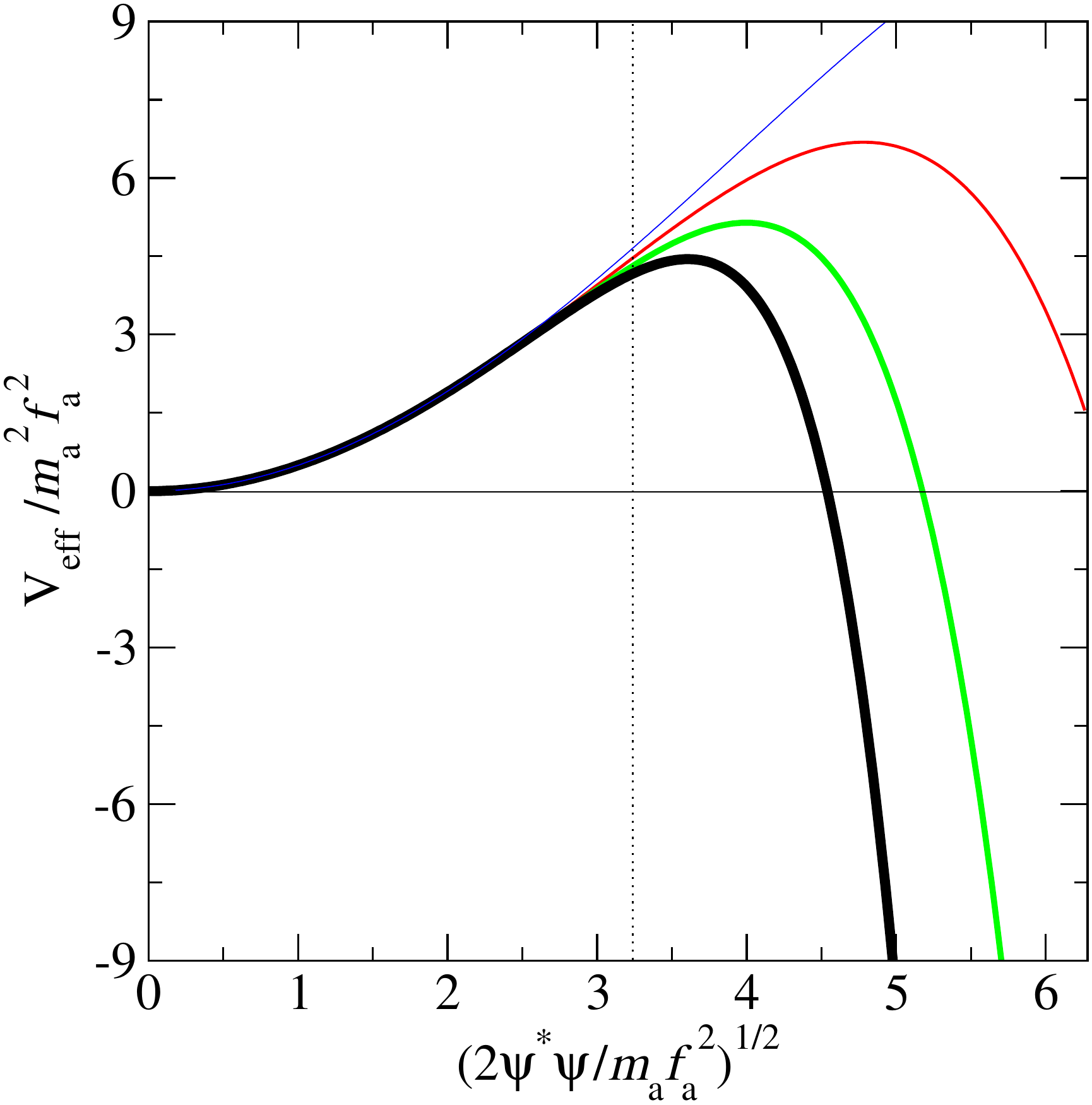}
\vspace*{0.0cm}
\caption{
Polynomial truncations of the effective potentials ${\cal V}_{\rm eff}$  as functions of $|\psi|$:
instanton effective potential (left panel)
and chiral effective potential with $z = 0.48$ (right panel).
The truncations of the potential after the 2nd, 3rd, 4th, and 5th powers of $\psi^*\psi$
are shown as successively thicker solid lines.
The vertical dotted line in the right panel
marks the radius of convergence of the chiral nonrelativistic reduction potential.
}
\label{fig:Veff5}
\end{figure}
%%%%%%%%%%%%%%%%%%%%%%%%%%%%%%%%%%%%%%%%%%%%%%%%

For the instanton potential,
the first four dimensionless coupling constants in Eq.~\eqref{v2345}
are $v_2 =-1$, $v_3 =-1.125$, $v_4 =-2.25$, and $v_5 =1.76$.
The coefficient $v_3$ has the opposite sign as $\lambda_6$.
The behavior of the first four polynomial truncations of the instanton effective  potential
seems to be compatible with an infinite radius of convergence.

For the chiral potential with $z=0.48$, the first four dimensionless coupling constants 
 in Eq.~\eqref{v2345}
are $v_2 = -0.343$, $v_3 = -0.376$, $v_4  = -1.841$, and $v_5  = -17.2$. 
These coefficients $v_n$ all have the same signs as $\lambda_{2n}$.
If this pattern continues, the first positive coefficient would be $v_9$.
The first four polynomial truncations of the chiral effective potential are shown in Fig.~\ref{fig:Veff5}.
The sequence of polynomial truncations converges rapidly for small values of  $\psi^*\psi$.
However the sequence seems to diverge for  $\psi^*\psi$ beyond
$r_c^2 m_a f_a^2/2$, where $r_c = 3.23$ is
the radius of convergence  of the chiral nonrelativistic reduction potential in Eq.~\eqref{r-conv}.

\subsection {Naive nonrelativistic reduction}
\label{sec:NRreduction}

A polynomial truncation of the power series in Eq.~\eqref{Veff-series}
for the effective potential ${\cal V}_{\rm eff}$
is useful only if $\psi^* \psi$ is much less than $m_af_a^2$.
Otherwise it is necessary to keep terms of all orders in $\psi^* \psi$.
A first approximation to the effective potential for axion EFT
that includes terms of all orders in $\psi^* \psi$
can be obtained by making a  naive nonrelativistic reduction.
The real field $\phi$ in the relativistic Hamiltonian density in Eq.~\eqref{H-phi} is replaced by
%-----------------
\begin{equation}
\phi(\bm{r},t)  = \frac{1}{\sqrt{2m_a} }
\left[ \psi(\bm{r},t) e^{-im_at} + \psi^*(\bm{r},t) e^{+im_at} \right],
\label{phi-psi}
\end{equation}
%----------------- 
where $\psi(\bm{r},t)$ is a  complex scalar field.
Most of the resulting terms in the Hamiltonian density have a rapidly oscillating phase factor $e^{ijm_at}$, 
where $j$ is a nonzero integer.
Upon dropping  terms with a rapidly oscillating phase factor 
and also dropping the terms proportional to $ \dot \psi$ and $ \dot \psi^*$, 
we obtain an effective Hamiltonian density  $\cal{H}$ of the form in Eq.~\eqref{Heff-psi}.
The resulting power series for the effective potential is
%-----------------
\begin{equation}
{\cal V}_{\rm eff}^{(0)}(\psi^* \psi)  = m_a \psi^* \psi + m_a^2 f_a^2 
\sum_{n=2}^{\infty} \frac{\lambda_{2n}}{(n!)^2} \left(\frac{\psi^* \psi}{2m_af_a^2} \right)^n.
\label{Vclass-series}
\end{equation}
%----------------- 
(The meaning of the superscript (0) will be clear later.)
In Eq.~\eqref{Vclass-series},
half of the mass term $ m_a \psi^* \psi$ comes from the $\dot{\phi}^2$ term in Eq.~\eqref{H-phi},
and the other half comes from the $\phi^2$ term in ${\cal V}(\phi)$.
We call the potential with this power series the {\it nonrelativistic reduction potential}.
By comparing the coefficients $v_n^{(0)} = \lambda_{2n}$ 
of the interaction terms in Eq.~\eqref{Vclass-series}
with the exact coefficients $v_n$ given in Eqs.~\eqref{v2345},
we see that only $v_2$ is correct.
Thus the nonrelativistic reduction potential has limited accuracy at small $\psi^* \psi$.

It is convenient to introduce a dimensionless  number density variable $\hat n$ defined by
%-----------------
\begin{equation}
\hat n = 2\psi^* \psi/(m_af_a^2).
\label{psi-hat}
\end{equation}
%----------------- 
By the ratio test, the radius of convergence in $\hat n$ 
of the power series for ${\cal V}_{\rm eff}^{(0)}$
is the same as the radius of convergence in $\phi^2/f_a^2$
of the power series for ${\cal V}$ in Eq.~\eqref{V-series}.
Inside the radius of convergence,
the nonrelativistic reduction potential ${\cal V}_{\rm eff}^{(0)}$ can be defined by the power series.
Outside the radius of convergence,
it is necessary to use a resummation method to calculate ${\cal V}_{\rm eff}^{(0)}$.
The nonrelativistic reduction potential represents a selective resummation of terms to all orders in $\psi^* \psi$.
It is likely to be a better approximation for $\psi^* \psi$ of order $m_af_a^2$
than the polynomial in $\psi^* \psi$ obtained by any truncation of the power series.

%%%%%%%%%%%%%%%%%%%%%%%%%%%%%%%%%%%%%%%%%%%%%%%%
\begin{figure}[t]
\centerline{ \includegraphics*[width=12cm,clip=true]{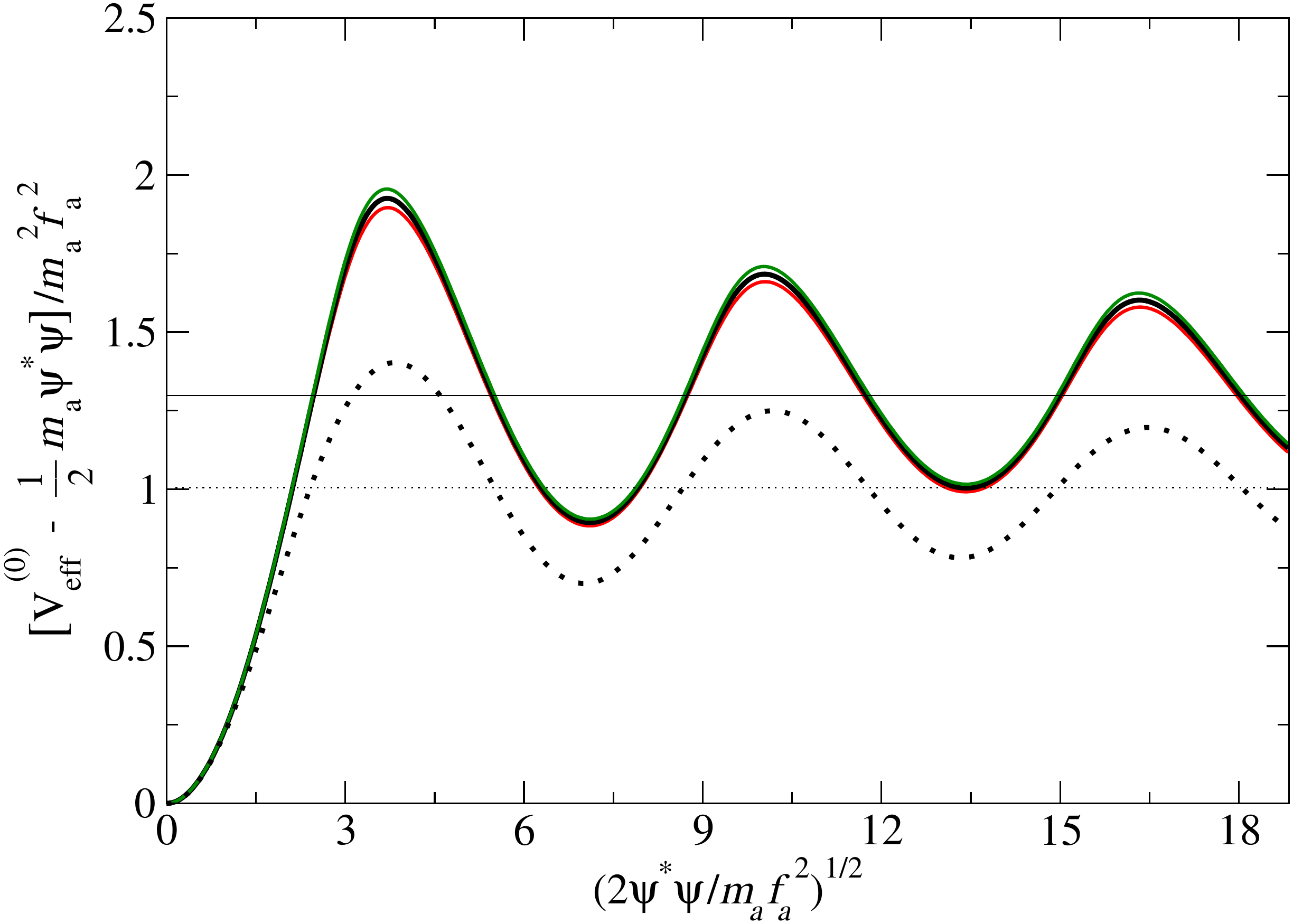} }
\vspace*{0.0cm}
\caption{Nonrelativistic reduction potentials ${\cal V}_{\rm eff}^{(0)}$
with $\tfrac12  m_a \psi^* \psi$ subtracted as functions of $|\psi|$:
instanton  potential (dotted curve)
and chiral potentials for $z=0.48$ (thicker solid curve)
and for $z=0.45$ and 0.51 (thinner solid curves).
The horizontal lines are the asymptotic values of the subtracted potentials at large $|\psi|$.
}
\label{fig:Vclass}
\end{figure}
%%%%%%%%%%%%%%%%%%%%%%%%%%%%%%%%%%%%%%%%%%%%%%%%

We first consider the {\it instanton nonrelativistic reduction potential}.
For the instanton potential in Eq.~\eqref{V-instanton}, 
the dimensionless coupling constants in Eq.~\eqref{Vclass-series} are $\lambda_{2n} = (-1)^{n+1}$.
The power series in Eq.~\eqref{Vclass-series} has an infinite radius of convergence
and it can be summed analytically:
%-----------------
\begin{equation}
{\cal V}_{\rm eff}^{(0)}(\psi^* \psi) =   \tfrac12  m_a \psi^* \psi
+ m_a^2 f_a^2 \left[ 1- J_0( \hat n^{1/2}) \right],
\label{Vclass-instanton}
\end{equation}
%----------------- 
where $J_0(z)$ is a Bessel function.  
This potential was first derived in Ref.~\cite{Eby:2014fya}.
The explicit term proportional to $\psi^* \psi$ is half the mass term in Eq.~\eqref{Vclass-instanton}.
The other half of the mass term comes from the  second term in 
Eq.~\eqref{Vclass-instanton}, which is a bounded function of $\psi^* \psi$.
The instanton nonrelativistic reduction potential is illustrated in Fig.~\ref{fig:Vclass}, 
where ${\cal V}_{\rm eff}^{(0)}$ 
with half the mass term subtracted is shown as a function of $\hat n^{1/2}$.
For small $\hat n$, this subtracted  potential is $\frac 12 m_a \psi^* \psi$.
For large $\hat n$, the subtracted  potential approaches $m_a^2 f_a^2$
with the asymptotic behavior 
%-----------------
\begin{equation}
{\cal V}_{\rm eff}^{(0)}(\psi^* \psi) -  \tfrac12  m_a \psi^* \psi \longrightarrow 
m_a^2 f_a^2 \left[ 1 - \left( \frac {4}{\pi^2 \hat n} \right)^{1/4} \cos(\hat n^{1/2} -\tfrac14 \pi) \right].
\label{largepsi-hat}
\end{equation}
%----------------- 
The damped oscillatory behavior at large $\hat n$ is a remnant
of the periodicity of the instanton potential.

We next consider the {\it chiral nonrelativistic reduction potential}.
If the chiral potential in Eq.~\eqref{V-chiral}  
is expanded as a cosine series as in Eq.~\eqref{V-cosine},
the nonrelativistic reduction potential can be expressed as a Bessel series.
To obtain the Bessel series, we expand $ \cos(j\phi/f_a)$ as a power series in $\phi$,
make the substitution in Eq.~\eqref{phi-psi}, and then drop rapidly oscillating terms.
The resulting power series in $\psi^* \psi$
can be summed analytically to get the Bessel function $J_0(j \hat n^{1/2})$.
The resulting nonrelativistic reduction potential is
%-----------------
\begin{equation}
{\cal V}_{\rm eff}^{(0)}(\psi^* \psi) =  \tfrac12  m_a \psi^* \psi
+m_a^2 f_a^2  \sum_{j=1}^\infty b_j \big[ J_0(j \hat n^{1/2}) -1 \big] .
\label{Vclass-Bessel}
\end{equation}
%----------------- 
The term $\frac12 m_a \psi^* \psi$ comes from the $\dot{\phi}^2$ term in Eq.~\eqref{H-phi},
and the second term comes from ${\cal V}(\phi)$ in Eq.~\eqref{V-cosine}.
The coefficients $b_j$ for the chiral potential are given in Eq.~\eqref{bj-chiral}.
Since the Bessel functions  $J_0(j \hat n^{1/2})$ are bounded functions of $\hat n$
and the coefficients $b_j$ decrease exponentially at large $j$, 
the sum over $j$ in Eq.~\eqref{Vclass-Bessel} converges.
The chiral nonrelativistic reduction potential seems to be the sum of 
$\frac12 m_a \psi^* \psi$ and a term that is a bounded function of $\psi^* \psi$.
The potential with $\frac12 m_a \psi^* \psi$ subtracted is shown in Fig.~\ref{fig:Vclass}.
It has the same qualitative behavior as the instanton nonrelativistic reduction potential.  
At large $\hat n$, the subtracted potential
approaches $1.30\,m_a^2 f_a^2$ for $z = 0.48$, 
oscillating around that value with an amplitude that decreases as $\hat n$ increases.
This oscillatory behavior at large $\hat n$ is a remnant
of the periodicity of the chiral potential.

The chiral nonrelativistic reduction potential has a power series expansion 
in $\psi^* \psi$ as in Eq.~\eqref{Vclass-series}.
The coefficient $\lambda_{2n}$ can be obtained analytically by expanding the 
chiral potential in Eq.~\eqref{V-chiral} in powers of $\phi$, 
and it is expressed as an infinite series in Eq.~\eqref{lambda2n-bj}.
The radius of convergence in $\hat n^{1/2}$ of the power series for ${\cal V}_{\rm eff}^{(0)}$,
which is given in Eq.~\eqref{r-conv}, is $r_c=3.23$ for $z = 0.48$.
Outside this radius of convergence, it is necessary to use a resummation method 
to calculate the nonrelativistic reduction potential ${\cal V}_{\rm eff}^{(0)}$.
The Bessel series for the  nonrelativistic reduction potential in Eq.~\eqref{Vclass-Bessel}
can be used to construct such a resummation method.
The Bessel function $J_0(j \hat n^{1/2})$ has a power series with an infinite radius of convergence.
The truncated Bessel expansion defined by truncating the sum over $j$ 
in Eq.~\eqref{Vclass-Bessel} after the $j_{\rm max}$ term 
therefore also has a power series with an infinite radius of convergence.
The coefficient $\lambda_{2n}(j_{\rm max})$ for that power series 
is obtained by truncating the series in Eq.~\eqref{lambda2n-bj} after the $j_{\rm max}$ term.
As $j_{\rm max}$ is increased, the truncated Bessel expansion converges 
 as a function of $j_{\rm max}$ to ${\cal V}_{\rm eff}^{(0)}$ out to increasingly larger values of $\hat n$.

\subsection{Improved effective potentials}
\label{sec:Semiclassical}

The nonrelativistic reduction potential ${\cal V}_{\rm eff}^{(0)}$ 
defined by the power series in Eq.~\eqref{Vclass-series}
is of limited accuracy at small $\psi^*\psi$, but it may provide a better approximation 
to the exact effective potential ${\cal V}_{\rm eff}$ for large $\psi^*\psi$ than
any truncation of the power series for ${\cal V}_{\rm eff}$.
The nonrelativistic reduction potential would be a more compelling approximation
if it was the first in a sequence of effective potentials.
We proceed to propose such a sequence 
that we call {\it improved effective potentials}.

%%%%%%%%%%%%%%%%%%%%%%%%%%%%%%%%%%%%%%%%%%%%%%%%
\begin{figure}[t]
\centerline{ \includegraphics*[width=12cm,clip=true]{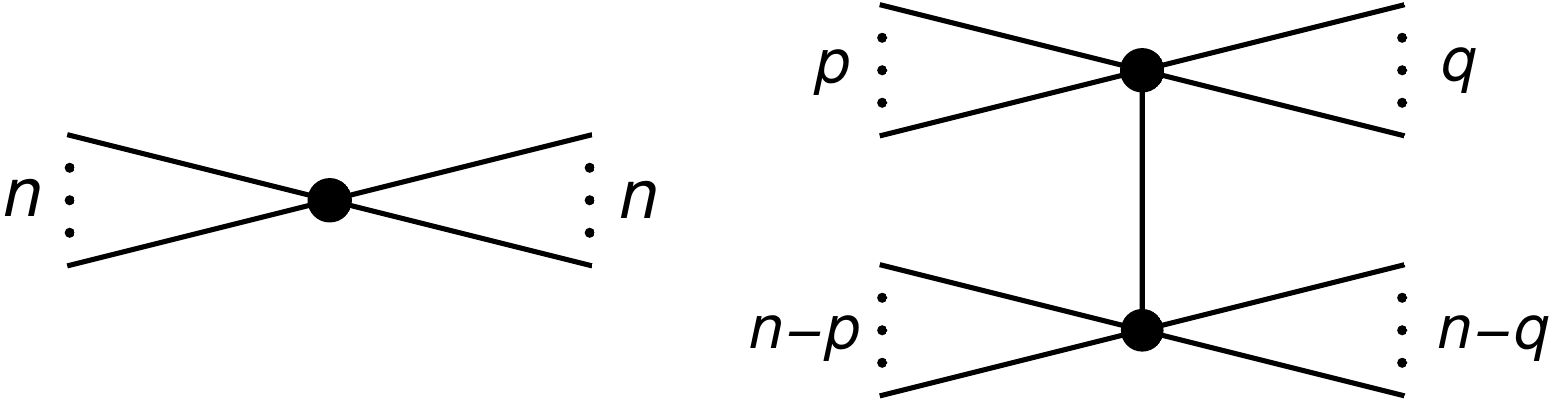} }
\vspace*{0.0cm}
\caption{The tree-level diagrams for $n \to n$ scattering in the relativistic axion theory
with no virtual axion lines (left diagram)  and with one virtual axion line (right diagram).
The corresponding diagrams  in axion EFT 
are all those with no virtual axion lines
and those with one virtual axion line for which $|p-q|=1$.
}
\label{fig:treen->n}
\end{figure}
%%%%%%%%%%%%%%%%%%%%%%%%%%%%%%%%%%%%%%%%%%%%%%%%

The power series in Eq.~\eqref{Vclass-series}  for the nonrelativistic reduction potential ${\cal V}_{\rm eff}^{(0)}$
can be obtained by approximating the coefficients $v_n$ in the power series  in Eq.~\eqref{Veff-series}
for the effective potential by $v_n^{(0)} \equiv \lambda_{2n}$.
In the effective field theory approach, this is equivalent to matching the contributions to 
the $n \to n$ scattering amplitude for all $n$ from diagrams with no virtual propagators.
These diagrams are the $2n$ axion vertex in the relativistic theory and the $n \to n$ vertex in axion EFT.
For $n=3$ and 4, these diagrams are the left-most diagram in Fig.~\ref{fig:3to3tree} 
and the top diagram in Fig.~\ref{fig:4to4tree}, respectively.
For general $n$, the set of diagrams is those in the left diagram in Fig.~\ref{fig:treen->n},
which has $n$ incoming lines attached to $n$ outgoing lines at a single vertex.  

A sequence ${\cal V}_{\rm eff}^{(k)}$, $k=0,1,2,\ldots$,
of effective potentials can be defined
by matching  the contributions to the $n \to n$ scattering amplitude for all $n$
from diagrams with at most $k$  virtual propagators.
The nonrelativistic reduction potential is the $k=0$ term in this sequence:
${\cal V}_{\rm eff}^{(0)}$.
We will refer to the potentials ${\cal V}_{\rm eff}^{(k)}$ with $k\ge1$ as 
{\it improved effective potentials}.
They will be defined by the power series expansion in Eq.~\eqref{Vclass-series}
with the coefficients $v_n$ replaced by $v_n^{(k)}$.
The coefficients $v_n^{(k)}$ agree with the exact coefficients $v_n$ for $n=2, \ldots,k+2$.
Thus the convergence of this sequence of effective potentials to ${\cal V}_{\rm eff}$
for small $\psi^* \psi$ is just as fast as the sequence of polynomial truncations of ${\cal V}_{\rm eff}$.
We expect the sequence ${\cal V}_{\rm eff}^{(k)}$ to be increasingly accurate 
for large $\psi^* \psi$ as $k$ increases, because increasingly larger classes of diagrams are summed.  
We expect the sequence ${\cal V}_{\rm eff}^{(k)}$ to converge to  ${\cal V}_{\rm eff}$
as $k \to \infty$, because all diagrams are included in this limit.

The first improved effective potential ${\cal V}_{\rm eff}^{(1)}$
is obtained by matching the contributions to scattering amplitudes 
from diagrams with zero or one virtual axion lines.
The set of diagrams for $n \to n$ scattering in the relativistic theory 
are shown in  Fig.~\ref{fig:treen->n}.
The left diagram has $n$ incoming lines attached to $n$ outgoing lines 
at a single vertex.  These are also diagrams in axion EFT.
The right diagram  in  Fig.~\ref{fig:treen->n} has two vertices connected by a single virtual axion line.
The first vertex has $p$ incoming lines and $q$ outgoing lines.
The corresponding diagrams in axion EFT are the subset for which $|p-q| = 1$.
The contributions to the dimensionless coupling constant $v_{n}$
from matching both sets of diagrams in Fig.~\ref{fig:treen->n} 
can be expressed as a sum over $p$ and $q$:
%-----------------
\begin{equation}
v_{n}^{(1)}  = \lambda_{2n} + \frac12
\sum_{p=0}^n {}' \sum_{q=0}^n {}' \binom{n}{p} \binom{n}{q} \lambda_{p+q+1}  \lambda_{2n-p-q+1}
\left( \frac{1-\delta_{|p-q|,1}}{(p-q)^2-1} - \frac{\delta_{|p-q|,1}}{4} \right).
\label{v3-lambda}
\end{equation}
%----------------- 
As indicated by the primes on the sums, the values of $p$ and $q$ are further constrained 
by the  requirement that $p+q$ be an odd integer ranging from 3 to $2n-3$.
This constraint can be made implicit by adopting  the conventions that $\lambda_2=0$
and that $ \lambda_m = 0$ if $m$ is odd.
The power series for the first improved effective potential ${\cal V}_{\rm eff}^{(1)}$ 
can be determined by inserting the coefficients $v_{n}^{(1)}$ in Eq.~\eqref{v3-lambda}
into the power series in Eq.~\eqref{Veff-series},
provided $\hat n$ is within the radius of convergence.
If $\hat n$ is inside the radius of convergence, the potential ${\cal V}_{\rm eff}^{(1)}$
can be defined by the power series.
If $\hat n$ is outside the radius of convergence,
it is necessary to use some resummation method to calculate ${\cal V}_{\rm eff}^{(1)}$.

%%%%%%%%%%%%%%%%%%%%%%%%%%%%%%%%%%%%%%%%%%%%%%%%
\begin{figure}[t]
\centerline{ \includegraphics*[width=12cm,clip=true]{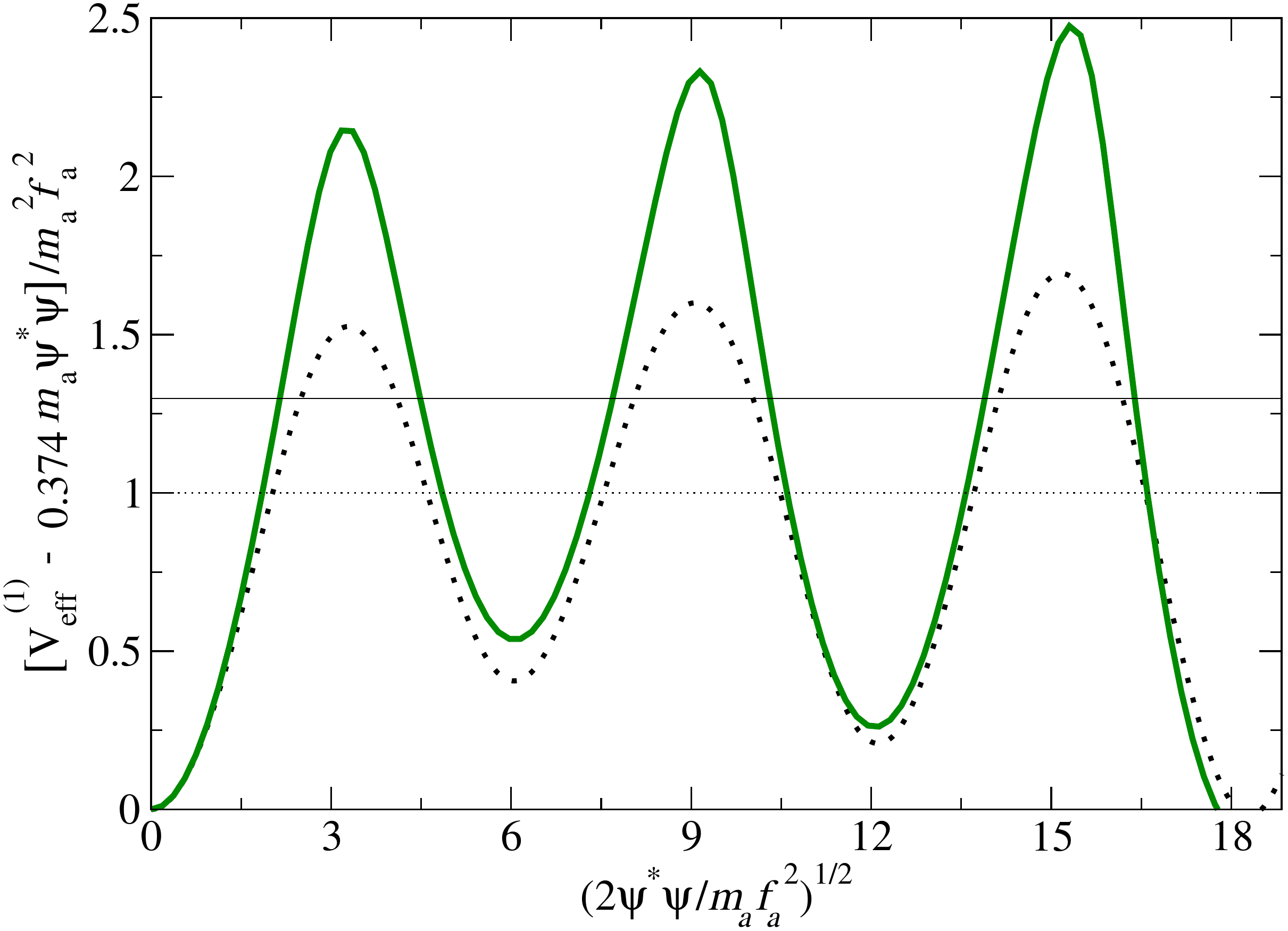} }
\vspace*{0.0cm}
\caption{First improved effective potentials ${\cal V}_{\rm eff}^{(1)}$
with $0.374\,  m_a \psi^* \psi$ subtracted as functions of $|\psi|$:
instanton potential (dotted curve) and chiral potential with $z=0.48$ (solid curve).
The horizontal lines are the asymptotic values of the 
corresponding subtracted nonrelativistic reduction potentials in Fig.~\ref{fig:Vclass}.
}
\label{fig:Vsemiclass}
\end{figure}
%%%%%%%%%%%%%%%%%%%%%%%%%%%%%%%%%%%%%%%%%%%%%%%%

We first consider the {\it first improved instanton effective potential}.
The power-series coefficients $v_n^{(1)}$ for ${\cal V}_{\rm eff}^{(1)}$ in Eq.~\eqref{v3-lambda}
are determined by the coefficients $\lambda_m$, which
are equal to $(-1)^{\frac12m+1}$ if $m$ is even and 0 if $m$ is odd.
The power series seems to have an infinite radius of convergence.
It seems to be the sum of a term proportional to $m_a\psi^* \psi$ 
with coefficient 0.374(2) and a term that  oscillates as a function of $\hat n^{1/2}$ at large $\hat n$.
In Fig.~\ref{fig:Vsemiclass},
we show the first improved instanton effective potential ${\cal V}_{\rm eff}^{(1)}$
with $0.374\,m_a\psi^* \psi$ subtracted as a function of $\hat n^{1/2}$.
At large $\hat n$, the subtracted potential seems to oscillate around the same value $m_a^2 f_a^2$ 
as the corresponding nonrelativistic reduction potential in Fig.~\ref{fig:Vclass},
except with a slowly increasing amplitude instead of a slowly decreasing amplitude.

We next consider the {\it first improved chiral effective potential}.
The power-series coefficients $v_n^{(1)}$ for ${\cal V}_{\rm eff}^{(1)}$ in Eq.~\eqref{v3-lambda}
are determined by the coefficients $\lambda_{2n}$
in the chiral nonrelativistic reduction potential ${\cal V}_{\rm eff}^{(0)}$.
The power series for ${\cal V}_{\rm eff}^{(1)}$ seems to have the same radius of convergence in  
$\hat n$
as the power series for ${\cal V}_{\rm eff}^{(0)}$,
whose radius of convergence is given in Eq.~\eqref{r-conv}.
The power series with coefficients $v_{n}^{(1)}$ in Eq.~\eqref{v3-lambda}
can be used to calculate ${\cal V}_{\rm eff}^{(1)}$ only for $\hat n$ inside the radius of convergence.
Outside that radius of convergence,
${\cal V}_{\rm eff}^{(1)}$  can be calculated using a resummation method
similar to that used for the nonrelativistic reduction potential.
If the Bessel series for ${\cal V}_{\rm eff}^{(0)}$ in Eq.~\eqref{Vclass-Bessel}
is truncated after the $j_{\rm max}$ term,
the coefficients $\lambda_{2n}(j_{\rm max})$ in its power series expansion in Eq.~\eqref{Veff-series}
are given by truncating the series in Eq.~\eqref{lambda2n-bj} after the $j_{\rm max}$ term.
Inserting the coefficients $\lambda_{2n}(j_{\rm max})$  into Eq.~\eqref{v3-lambda}
defines coefficients $v_{n}^{(1)}(j_{\rm max})$ that depend on $j_{\rm max}$.
The power series obtained by inserting these coefficients 
$v_{n}^{(1)}(j_{\rm max})$ into Eq.~\eqref{Veff-series} 
seems to have an infinite radius of convergence.  As $j_{\rm max}$ increases, 
the power series converges as a function of $j_{\rm max}$ out to larger and larger values of $\hat n$.
We identify the function to which it converges 
as the first improved chiral effective potential ${\cal V}_{\rm eff}^{(1)}$.

We have used this resummation method to calculate 
the first improved chiral effective potential ${\cal V}_{\rm eff}^{(1)}$  with $z=0.48$.
It seems to be the sum of a term proportional to $m_a\psi^* \psi$ 
with coefficient 0.374(9) and a term that  oscillates as a function of $\hat n^{1/2}$ at large $\hat n$.
The coefficient of the $m_a\psi^* \psi$ term seems to have the same value 0.374 
as for the first improved instanton effective potential.
In Fig.~\ref{fig:Vsemiclass},
we show the first improved chiral effective potential ${\cal V}_{\rm eff}^{(1)}$
with $0.374\,m_a\psi^* \psi$ subtracted as a function of $\hat n^{1/2}$.
At large $\hat n$,  the subtracted potential seems to oscillate around the same value 
$1.30\,m_a^2 f_a^2$ as the chiral nonrelativistic reduction potential in Fig.~\ref{fig:Vclass},
except with a slowly increasing amplitude instead of a slowly decreasing amplitude.

It is evident from the comparison of Figs.~\ref{fig:Vsemiclass} and \ref{fig:Vclass} 
that there are significant differences between the first improved effective potential 
and the corresponding nonrelativistic reduction potential at large $\psi$.
This raises the question whether the sequence of improved effective potentials
${\cal V}_{\rm eff}^{(k)}$ does in fact converge as $k \to \infty$.
It would be worthwhile to calculate the second improved effective potential 
${\cal V}_{\rm eff}^{(2)}$ to see whether there is any sign of convergence 
of the sequence ${\cal V}_{\rm eff}^{(k)}$.
If the sequence does not converge, it will be necessary to develop a better way 
to calculate the effective potential ${\cal V}_{\rm eff}(\psi^*\psi)$ for axion EFT
at large values of $\psi^*\psi$.

%\newpage

\section{Summary}
\label{sec:Summary}

Axions can be described by a relativistic quantum field theory with a real scalar field $\phi(x)$.
The self-interaction potential ${\cal V}(\phi)$ for axions is a periodic function of $\phi$.
Most phenomenological investigations of axions have been carried out using the 
{\it instanton potential} in  Eq.~\eqref{V-instanton},
but a more accurate potential for the QCD axion is the {\it chiral potential} in Eq.~\eqref{V-chiral},
which depends on $z = m_u/m_d$.
The two potentials are compared in Fig.~\ref{fig:potential}.
There are quantitative differences between the two potentials.
There is also an important qualitative difference that is illustrated in Fig.~\ref{fig:V5}.
The power series in $\phi$ for the instanton potential has an infinite radius of convergence,
while the power series for the chiral potential has a radius of convergence in $\phi$ 
of $r_c f_a$, where $r_c$ is given in Eq.~\eqref{r-conv}.

Nonrelativistic axions can be described more simply by a nonrelativistic effective field theory 
called {\it axion EFT} with a complex scalar field $\psi(\bm{r},t)$.
In axion EFT, the self-interactions of axions are described by an effective potential 
${\cal V}_{\rm eff}(\psi^*\psi)$ that has no simple periodicity properties.
We have calculated ${\cal V}_{\rm eff}$ to 5th order in $\psi^*\psi$
by matching low-energy scattering amplitudes in the relativistic theory and in axion EFT.
The coefficients of the powers of $\psi^*\psi$ are given in Eqs.~\eqref{v2345}
in terms of the coefficients in the expansion of ${\cal V}(\phi)$ in powers of $\phi^2$.
The first four polynomial truncations of the {\it instanton effective potential} 
and the {\it chiral effective potential} are shown in Fig.~\ref{fig:Veff5}.
There is a qualitative difference between the apparent convergence of the 
two effective potentials.
The instanton effective potential seems to have an infinite radius of convergence,
while the chiral effective potential seems to have an finite radius of convergence
in $\psi^*\psi$ that is roughly $r_c^2 m_a f_a^2/2$.

If the number density $\psi^*\psi$ is too large, 
the effective potential ${\cal V}_{\rm eff}$ cannot be 
approximated by a truncation of its expansion in powers of $\psi^*\psi$.
In Sec.~\ref{sec:Semiclassical}, we introduced a sequence of {\it improved effective potentials}
${\cal V}_{\rm eff}^{(k)}$ that  resum terms of all orders in $\psi^*\psi$.
The sequence is defined diagrammatically in terms of the matching procedure for axion EFT.
The coefficients in the expansion of ${\cal V}_{\rm eff}^{(k)}$ in powers of $\psi^*\psi$
are exact through order $k+2$.
In the limit $k \to \infty$, all diagrams are included,
so it is plausible that ${\cal V}_{\rm eff}^{(k)}$  converges to ${\cal V}_{\rm eff}$ in this limit.

We call the $k=0$ potential ${\cal V}_{\rm eff}^{(0)}$ 
in the sequence of systematically improvable effective potentials
the {\it nonrelativistic reduction potential}, and we denote it also by ${\cal V}_{\rm eff}^{(0)}$.
It can be defined by the coefficients in its power series in Eq.~\eqref{Vclass-series}, 
which are determined by the power-series coefficients for ${\cal V}(\phi)$. 
The instanton nonrelativistic reduction potential, which is given in Eq.~\eqref{Vclass-instanton},
was first derived in Ref.~\cite{Eby:2014fya}.
The instanton nonrelativistic reduction potential and the chiral nonrelativistic reduction potential
are compared in Fig.~\ref{fig:Vclass}.
The instanton nonrelativistic reduction potential is the sum of $\frac12 m_a \psi^*\psi$
and a term that at large $|\psi|$ oscillates as a function of $\hat n^{1/2}$ around $m_a^2 f_a^2$ 
with a decreasing amplitude.
The chiral nonrelativistic reduction potential for $z = 0.48$ is the sum of $\frac12 m_a \psi^*\psi$
and a term that at large $|\psi|$ seems to oscillate around $1.3\,m_a^2 f_a^2$ 
with a decreasing amplitude.
There is a qualitative difference between the two nonrelativistic reduction potentials 
in the convergence properties of their power series in $\psi^*\psi$.
The instanton nonrelativistic reduction potential has an infinite radius of convergence, 
while the radius of convergence in $\psi^*\psi$ 
of the power series for the chiral nonrelativistic reduction potential is $r_c^2 m_a f_a^2/2$.
We introduced a resummation method 
based on the cosine expansion of the relativistic chiral potential 
to calculate the chiral nonrelativistic reduction potential for larger values of $\psi^*\psi$.

We call the $k=1$ potential ${\cal V}_{\rm eff}^{(1)}$
in the sequence of systematically improvable effective potentials
the {\it first improved effective potential}.
It can be defined by the coefficients in its power series, which are given in 
Eq.~\eqref{v3-lambda} in terms of the power-series coefficients for ${\cal V}(\phi)$. 
The first improved instanton effective potential and the 
first improved chiral effective potential with $z = 0.48$ are compared in Fig.~\ref{fig:Vsemiclass}.
They both seem to be the sum of a term proportional to $m_a \psi^*\psi$
with a coefficient that is approximately 0.374
and a term that at large $|\psi|$ oscillates as a function of $\hat n^{1/2}$ with an increasing amplitude.
The second term seems to oscillate around the same value as for the 
corresponding nonrelativistic reduction potential,
which is $m_a^2 f_a^2$ for the instanton potential 
and $1.3\,m_a^2 f_a^2$ for the chiral potential with $z = 0.48$.
There seems to be a qualitative difference between these two improved effective potentials 
in the convergence properties of their power series in $\psi^*\psi$.
The first improved instanton effective potential seems to have an infinite radius of convergence, 
while the radius of convergence in $\psi^*\psi$ of the power series  
for the first improved chiral effective potential seems to be roughly $r_c^2 m_a f_a^2/2$.
We used a resummation method based on the cosine series for the relativistic chiral potential
to calculate the first improved chiral effective potential for larger values of $\psi^*\psi$.

The differences between the nonrelativistic reduction potential
and the first improved effective potential
can be seen by comparing Figs.~\ref{fig:Vclass} and \ref{fig:Vsemiclass}.
There is an important quantitative difference in the terms proportional to $m_a \psi^* \psi$,
which have been subtracted in these two figures.
Its coefficient is $\frac12$ for the nonrelativistic reduction potentials
and 0.374 for the first improved effective potentials.
There is an important qualitative difference in the behavior at large  $\psi^* \psi$,
where the nonrelativistic reduction potentials oscillate with decreasing amplitudes and
the first improved effective potentials  oscillate with increasing amplitudes.
These qualitative and quantitative differences raise questions about the convergence 
of the sequence ${\cal V}_{\rm eff}^{(k)}$ of effective potentials.
Calculations of the next potential ${\cal V}_{\rm eff}^{(2)}$ in the sequence 
could shed some light on this issue.  If the sequence ${\cal V}_{\rm eff}^{(k)}$ does not converge, 
it will be necessary to develop a better way to calculate the effective potential 
${\cal V}_{\rm eff}(\psi^*\psi)$ for axion EFT at large values of $\psi^*\psi$.

An important application of axion EFT is to Bose-Einstein condensates of axions.
The effective potential ${\cal V}_{\rm eff}(\psi^* \psi)$ is the mean-field energy of a condensate
in which the quantum field has expectation value $\psi$.
Sikivie and Yang have argued that the dark matter halo of a galaxy 
is a dilute Bose-Einstein condensate (BEC)  of axions \cite{Sikivie:2009qn}.
The $(\psi^* \psi)^2$ term in the effective potential is relevant to the thermalization of axions,
although gravitational interactions provide a more effective
thermalization mechanism \cite{Sikivie:2009qn,Erken:2011dz}.
The negative sign of the coefficient of the $(\psi^* \psi)^2$ term implies that a homogeneous BEC
is unstable to fluctuations that increase the local density.
This instability may limit the coherence length of an axion BEC
to regions much smaller than a galaxy \cite{Guth:2014hsa}.

Axion stars are gravitationally bound collections of axions  \cite{Tkachev:1991ka}.
The axions can be described by the real scalar field $\phi(x)$ of a relativistic field theory, 
with self-interactions given by the relativistic potential ${\cal V}(\phi)$
and with gravitational interactions given by general relativity.
Approximate solutions of the resulting equations for stable axion stars  
were first calculated numerically
by Barranco and Bernal \cite{Barranco:2010ib}.
Solutions were found only for axion stars with mass $M$ below a critical value $M_*$ 
that was determined numerically.  
We refer to these solutions as {\it dilute axions stars},
because the number density of axions is much less than $m_a f_a^2$, even at the center of the star.
In the dilute axion star, the attractive forces from gravity and from axion pair interactions 
are balanced by the kinetic pressure of the axions.
The axions in a dilute axion star are nonrelativistic,
so they can be described accurately and more simply by the complex scalar field $\psi({\bm r}, t)$
of axion EFT,  
with self-interactions given by the effective potential ${\cal V}_{\rm eff}(\psi^* \psi)$
and with gravitational interactions given by Newtonian gravity.
Because the axions are dilute, 
the effective potential ${\cal V}_{\rm eff}$ can be truncated after the $(\psi^* \psi)^2$ term.
Accurate numerical solutions of the resulting equations were calculated 
by Chavanis and Delfini  \cite{Chavanis:2011zm}.
The critical mass above which a dilute axion star is unstable to collapse is
%-----------------
\begin{equation}
M_* = 10.1\, |\lambda_4|^{-1/2}\big(\hbar f_a^2/ G m_a^2 c^3 \big)^{1/2},
\label{Mmax-axion}
\end{equation}
%----------------- 
where $G$ is Newton's gravitational constant.
If the axion mass is $m_a = 10^{-4\pm 1}$~eV and if $\lambda_4 = -0.343$, 
this critical mass is $10^{-13\mp 2}\,M_\odot$, where $M_\odot$ is the mass of the sun.
The critical mass  is comparable to the mass of an asteroid.

If the mass of a dilute axion star exceeds the critical mass $M_*$ in Eq.~\eqref{Mmax-axion},
the star is unstable to collapse.
As the axion star implodes, the number density at its center increases.
When $\psi^* \psi$  becomes comparable to $m_a f_a^2$,
terms in the effective potential of all orders in $\psi^* \psi$ become important.
In this case, it is necessary to use an approximation to ${\cal V}_{\rm eff}$
that includes term of all orders, such as one of the improved effective potentials
${\cal V}_{\rm eff}^{(k)}$ defined in Sec.~\ref{sec:Semiclassical}.
It may be possible to describe the collapse of the dilute axion star by solving the 
time-dependent field equations of axion EFT.

One possibility for the remnant from the collapse of a dilute axion star is a
{\it dense axion star}, in which the attractive force from gravity 
is balanced by the mean-field  pressure of the axion Bose-Einstein condensate~\cite{Braaten:2015eeu}.
In a {\it dense axion star}, the number density of axions 
has values larger than $m_a f_a^2$ in the interior of the star,
so the effective potential ${\cal V}_{\rm eff}$ cannot be approximated 
by a truncation of its power series in $\psi^* \psi$.
In Ref.~\cite{Braaten:2015eeu}, the mass-radius relation 
for dense axion stars was calculated under the assumption that axion self-interactions 
are described by the instanton nonrelativistic reduction potential.
It would be worthwhile to calculate the mass-radius relation 
and other properties of dense axion stars using the chiral nonrelativistic reduction potential, 
which may be more accurate.
It would also be worthwhile to calculate these properties
using the first improved effective potential to see how strongly they depend 
on the approximation for the effective potential of axion EFT.

Axion EFT should  be useful to address many of the important theoretical issues 
concerning axion dark matter.
Do gravitational interactions provide a sufficiently effective thermalization mechanism 
that axions remain in a Bose-Einstein condensate, 
as argued in Refs.~\cite{Sikivie:2009qn,Erken:2011dz}?
Is the coherence of the axion Bose-Einstein condensate limited by instabilities 
to regions the size of an asteroid, as suggested in Ref.~\cite{Guth:2014hsa}?
What is the fate of a dilute axion star if it accretes enough axions
so that it exceeds the critical mass $M_*$ in Eq.~\eqref{Mmax-axion}
and begins to collapse?
The answers to these questions are important for determining
whether the QCD axion remains a viable candidate for the dark matter particle.

\begin{acknowledgments}
This research was supported in part by the
Department of Energy under grant DE- SC0011726
and by the National Science Foundation under grant PHY-1310862.
We thank D.H.~Smith and F.~Werner for useful comments on
the convergence of power series.
\end{acknowledgments}

%\newpage


\begin{thebibliography}{99}

%\cite{Peccei:1977hh}
\bibitem{Peccei:1977hh} 
  R.D.~Peccei and H.R.~Quinn,
$CP$ conservation in the presence of instantons,
  Phys.\ Rev.\ Lett.\  {\bf 38}, 1440 (1977).
%  doi:10.1103/PhysRevLett.38.1440
  %%CITATION = doi:10.1103/PhysRevLett.38.1440;%%

%\cite{Weinberg:1977ma}
\bibitem{Weinberg:1977ma}
  S.~Weinberg,
A new light boson?,
  Phys.\ Rev.\ Lett.\  {\bf 40}, 223 (1978).
%  doi:10.1103/PhysRevLett.40.223
  %%CITATION = doi:10.1103/PhysRevLett.40.223;%%

%\cite{Wilczek:1977pj}
\bibitem{Wilczek:1977pj}
  F.~Wilczek,
Problem of strong $P$ and $T$ invariance in the presence of instantons,
  Phys.\ Rev.\ Lett.\  {\bf 40}, 279 (1978).
%  doi:10.1103/PhysRevLett.40.279
  %%CITATION = doi:10.1103/PhysRevLett.40.279;%%

%\cite{Kim:2008hd}
\bibitem{Kim:2008hd} 
  J.~E.~Kim and G.~Carosi,
Axions and the strong $CP$ problem,
  Rev.\ Mod.\ Phys.\  {\bf 82}, 557 (2010)
%  doi:10.1103/RevModPhys.82.557
  [arXiv:0807.3125].
  %%CITATION = doi:10.1103/RevModPhys.82.557;%%
  
%\cite{Davis:1986xc}
\bibitem{Davis:1986xc} 
  R.L.~Davis,
Cosmic axions from cosmic strings,
  Phys.\ Lett.\ B {\bf 180}, 225 (1986).
%  doi:10.1016/0370-2693(86)90300-X
  %%CITATION = doi:10.1016/0370-2693(86)90300-X;%%
  
%\cite{Harari:1987ht}
\bibitem{Harari:1987ht} 
  D.~Harari and P.~Sikivie,
On the evolution of global strings in the early universe,
  Phys.\ Lett.\ B {\bf 195}, 361 (1987).
%  doi:10.1016/0370-2693(87)90032-3
  %%CITATION = doi:10.1016/0370-2693(87)90032-3;%%
  
%\cite{Preskill:1982cy}
\bibitem{Preskill:1982cy} 
  J.~Preskill, M.B.~Wise, and F.~Wilczek,
Cosmology of the invisible axion,
  Phys.\ Lett.\ B {\bf 120}, 127 (1983).
%  doi:10.1016/0370-2693(83)90637-8
  %%CITATION = doi:10.1016/0370-2693(83)90637-8;%%
  
%\cite{Abbott:1982af}
\bibitem{Abbott:1982af} 
  L.F.~Abbott and P.~Sikivie,
A cosmological bound on the invisible axion,
  Phys.\ Lett.\ B {\bf 120}, 133 (1983).
%  doi:10.1016/0370-2693(83)90638-X
  %%CITATION = doi:10.1016/0370-2693(83)90638-X;%%

%\cite{Dine:1982ah}
\bibitem{Dine:1982ah} 
  M.~Dine and W.~Fischler,
The not so harmless axion,
  Phys.\ Lett.\ B {\bf 120}, 137 (1983).
%  doi:10.1016/0370-2693(83)90639-1
  %%CITATION = doi:10.1016/0370-2693(83)90639-1;%%
  
%\cite{Sikivie:2009qn}
\bibitem{Sikivie:2009qn} 
  P.~Sikivie and Q.~Yang,
Bose-Einstein condensation of dark matter axions,
  Phys.\ Rev.\ Lett.\  {\bf 103}, 111301 (2009)
%  doi:10.1103/PhysRevLett.103.111301
  [arXiv:0901.1106].
  %%CITATION = ARXIV:0901.1106;%%
  
%\cite{Erken:2011dz}
\bibitem{Erken:2011dz} 
  O.~Erken, P.~Sikivie, H.~Tam, and Q.~Yang,
Cosmic axion thermalization,
  Phys.\ Rev.\ D {\bf 85}, 063520 (2012)
%  doi:10.1103/PhysRevD.85.063520
  [arXiv:1111.1157].
  %%CITATION = doi:10.1103/PhysRevD.85.063520;%%
  
%\cite{Saikawa:2012uk}
\bibitem{Saikawa:2012uk} 
  K.~Saikawa and M.~Yamaguchi,
Evolution and thermalization of dark matter axions in the condensed regime,
  Phys.\ Rev.\ D {\bf 87}, 085010 (2013)
%  doi:10.1103/PhysRevD.87.085010
%  [arXiv:1210.7080 [hep-ph]].
  [arXiv:1210.7080].
  %%CITATION = doi:10.1103/PhysRevD.87.085010;%%
  
%\cite{Davidson:2013aba}
\bibitem{Davidson:2013aba} 
  S.~Davidson and M.~Elmer,
Bose-Einstein condensation of the classical axion field in cosmology?,
  JCAP {\bf 1312}, 034 (2013)
%  doi:10.1088/1475-7516/2013/12/034
  [arXiv:1307.8024].
  %%CITATION = doi:10.1088/1475-7516/2013/12/034;%%

%\cite{Noumi:2013zga}
\bibitem{Noumi:2013zga} 
  T.~Noumi, K.~Saikawa, R.~Sato, and M.~Yamaguchi,
Effective gravitational interactions of dark matter axions,
  Phys.\ Rev.\ D {\bf 89}, 065012 (2014)
%  doi:10.1103/PhysRevD.89.065012
%  [arXiv:1310.0167 [hep-ph]].
  [arXiv:1310.0167].
  %%CITATION = doi:10.1103/PhysRevD.89.065012;%%

%\cite{Davidson:2014hfa}
\bibitem{Davidson:2014hfa} 
  S.~Davidson,
Axions: Bose-Einstein condensate or classical field?,
  Astropart.\ Phys.\  {\bf 65}, 101 (2015)
%  doi:10.1016/j.astropartphys.2014.12.007
%  [arXiv:1405.1139 [hep-ph]].
  [arXiv:1405.1139].
  %%CITATION = doi:10.1016/j.astropartphys.2014.12.007;%%

%\cite{Braaten:2015eeu}
\bibitem{Braaten:2015eeu} 
  E.~Braaten, A.~Mohapatra and H.~Zhang,
Dense axion stars,
 Phys.\ Rev.\ Lett.\  {\bf 117}, 121801 (2016)
  [arXiv:1512.00108].
%  arXiv:1512.00108 [hep-ph].
  %%CITATION = ARXIV:1512.00108;%%
  
%\cite{Peccei:1977ur}
\bibitem{Peccei:1977ur} 
  R.D.~Peccei and H.R.~Quinn,
Constraints imposed by $CP$ conservation in the presence of instantons,
  Phys.\ Rev.\ D {\bf 16}, 1791 (1977).
%  doi:10.1103/PhysRevD.16.1791
  %%CITATION = doi:10.1103/PhysRevD.16.1791;%%
  
 %\cite{DiVecchia:1980yfw}
\bibitem{DiVecchia:1980yfw} 
  P.~Di Vecchia and G.~Veneziano,
Chiral dynamics in the large $N$ limit,
  Nucl.\ Phys.\ B {\bf 171}, 253 (1980).
%  doi:10.1016/0550-3213(80)90370-3
  %%CITATION = doi:10.1016/0550-3213(80)90370-3;%%
 
%\cite{diCortona:2015ldu}
\bibitem{diCortona:2015ldu} 
  G.G.~di Cortona, E.~Hardy, J.P.~Vega and G.~Villadoro,
The QCD axion, precisely,
  JHEP {\bf 1601}, 034 (2016)
%  doi:10.1007/JHEP01(2016)034
  [arXiv:1511.02867].
  %%CITATION = doi:10.1007/JHEP01(2016)034;%%
  
%\cite{Kim:1979if}
\bibitem{Kim:1979if} 
  J.E.~Kim,
Weak interaction singlet and strong $CP$ invariance,
  Phys.\ Rev.\ Lett.\  {\bf 43}, 103 (1979).
%  doi:10.1103/PhysRevLett.43.103
  %%CITATION = doi:10.1103/PhysRevLett.43.103;%%

%\cite{Shifman:1979if}
\bibitem{Shifman:1979if} 
  M.A.~Shifman, A.I.~Vainshtein, and V.I.~Zakharov,
Can confinement ensure natural $CP$ invariance of strong interactions?,
  Nucl.\ Phys.\ B {\bf 166}, 493 (1980).
%  doi:10.1016/0550-3213(80)90209-6
  %%CITATION = doi:10.1016/0550-3213(80)90209-6;%%
  
%\cite{Dine:1981rt}
\bibitem{Dine:1981rt} 
  M.~Dine, W.~Fischler, and M.~Srednicki,
A simple solution to the strong $CP$ problem with a harmless axion,
  Phys.\ Lett.\ B {\bf 104}, 199 (1981).
%  doi:10.1016/0370-2693(81)90590-6
  %%CITATION = doi:10.1016/0370-2693(81)90590-6;%%

%\cite{Zhitnitsky:1980tq}
\bibitem{Zhitnitsky:1980tq} 
 A.R.~Zhitnitsky,
On possible suppression of the axion hadron interactions,
  Sov.\ J.\ Nucl.\ Phys.\  {\bf 31}, 260 (1980)
  [Yad.\ Fiz.\  {\bf 31}, 497 (1980)].
  %%CITATION = SJNCA,31,260;%%

\bibitem{Watson:1918}
G.N.~Watson,
Asymptotic expansions of hypergeometric functions,
Trans.\ Cambridge Philos.\ Soc.\ {\bf 22}, 277 (1918).

%\cite{Eby:2014fya}
\bibitem{Eby:2014fya} 
  J.~Eby, P.~Suranyi, C.~Vaz, and L.C.R.~Wijewardhana,
Axion stars in the infrared limit,
   JHEP {\bf 1503}, 080 (2015)
% doi:10.1007/JHEP03(2015)080
  [arXiv:1412.3430].
  %%CITATION = ARXIV:1412.3430;%%
  
%\cite{Tkachev:1991ka}
\bibitem{Tkachev:1991ka} 
  I.I.~Tkachev,
On the possibility of Bose star formation,
  Phys.\ Lett.\ B {\bf 261}, 289 (1991).
%  doi:10.1016/0370-2693(84)90764-0
  %%CITATION = doi:10.1016/0370-2693(91)90330-S;%%

%\cite{Barranco:2010ib}
\bibitem{Barranco:2010ib} 
  J.~Barranco and A.~Bernal,
Self-gravitating system made of axions,
  Phys.\ Rev.\ D {\bf 83}, 043525 (2011)
%  doi:10.1103/PhysRevD.83.043525
  [arXiv:1001.1769].
  %%CITATION = doi:10.1103/PhysRevD.83.043525;%%
    
 %\cite{Chavanis:2011zm}
\bibitem{Chavanis:2011zm} 
  P.H.~Chavanis and L.~Delfini,
Mass-radius relation of Newtonian self-gravitating Bose-Einstein condensates with short-range interactions: 
II. Numerical results,
  Phys.\ Rev.\ D {\bf 84}, 043532 (2011)
%  doi:10.1103/PhysRevD.84.043532
  [arXiv:1103.2054].
  %%CITATION = doi:10.1103/PhysRevD.84.043532;%%   

%\cite{Guth:2014hsa}
\bibitem{Guth:2014hsa} 
  A.H.~Guth, M.P.~Hertzberg, and C.~Prescod-Weinstein,
Do dark matter axions form a condensate with long-range correlation?,
  Phys.\ Rev.\ D {\bf 92}, 103513 (2015)
%  doi:10.1103/PhysRevD.92.103513
  [arXiv:1412.5930].


\end{thebibliography}
\end{document}